\documentclass{aa} 
\DeclareUnicodeCharacter{0327}{}

\usepackage{natbib,twoopt}
\usepackage[colorlinks=true,citecolor=blue,linkcolor=blue,urlcolor=blue,breaklinks=true]{hyperref} 
\bibpunct{(}{)}{;}{a}{}{,}             
\makeatletter
  \newcommandtwoopt{\citeads}[3][][]{\href{http://adsabs.harvard.edu/abs/#3}%
    {\def\hyper@linkstart##1##2{}%
     \let\hyper@linkend\@empty\citealp[#1][#2]{#3}}}
  \newcommandtwoopt{\citepads}[3][][]{\href{http://adsabs.harvard.edu/abs/#3}%
    {\def\hyper@linkstart##1##2{}%
     \let\hyper@linkend\@empty\citep[#1][#2]{#3}}}
  \newcommandtwoopt{\citetads}[3][][]{\href{http://adsabs.harvard.edu/abs/#3}%
    {\def\hyper@linkstart##1##2{}%
     \let\hyper@linkend\@empty\citet[#1][#2]{#3}}}
  \newcommandtwoopt{\citeyearads}[3][][]%
    {\href{http://adsabs.harvard.edu/abs/#3}
    {\def\hyper@linkstart##1##2{}%
     \let\hyper@linkend\@empty\citeyear[#1][#2]{#3}}}
\makeatother

\usepackage{graphicx}
\usepackage{multirow}
\usepackage{txfonts}

\newcommand{\mout}{$\dot{m}_\text{ejec}$}
\newcommand{\afunnel}{$\theta_f$}
\newcommand{\msun}{$M_\odot$}

\newcommand{\nustar}{\textit{NuSTAR}}
\newcommand{\xmm}{\textit{XMM--Newton}}
\newcommand{\chandra}{\textit{Chandra}}
\newcommand{\swift}{\textit{Swift}}
\newcommand{\hubble}{\textit{Hubble Space Telescope}}
\newcommand{\hst}{\textit{HST}}
\newcommand{\epicpn}{EPIC-\textit{pn}}

\newcommand{\tbabs}{\texttt{tbabs}}

\newcommand{\cflux}{\texttt{cflux}}

\newcommand{\xspec}{\textsc{XSPEC}}

\newcommand{\chisq}{$\chi^2$}
\newcommand{\chisqr}{$\chi^2_r$}
\newcommand{\nh}{$n_{\text{H}}$}

\newcommand{\errors}[2]{$^{+#1}_{-#2}$}
\newcommand{\ls}{Lomb--Scargle}

\newcommand{\ho}{Holmberg~II X--1}
\newcommand{\ngc}{NGC 5204~X--1}

\begin{document} 

   \title{Discovery of a recurrent spectral evolutionary cycle in the ultra-luminous X-ray sources \ho\ and \ngc\ }
\titlerunning{Evolutionary cycles in the ULXs \ho\ and \ngc}

   \author{A. G\'urpide
          \inst{1}, O. Godet\inst{1},
          G. Vasilopoulos\inst{2,3},
          N. A. Webb\inst{1},
          J.-F. Olive\inst{1}
          }
          \authorrunning{A. G\'urpide et al.} 

   \institute{IRAP, Université de Toulouse, CNRS, CNES, 9 avenue du Colonel Roche, 31028, Toulouse, France
   \and 
   Department of Astronomy, Yale University, PO Box 208101, New Haven, CT 06520-8101, USA
   \and
   Université de Strasbourg, CNRS, Observatoire astronomique de Strasbourg, UMR 7550, 67000, Strasbourg, France}
    

   \date{}

 
  \abstract
   {Most ultra-luminous X-ray sources (ULXs) are now thought to be powered by stellar-mass compact objects accreting at super-Eddington rates. While the discovery of evolutionary cycles have marked a breakthrough in our understanding of the accretion flow changes in the sub-Eddington regime in Galactic black hole binaries, their evidence in the super-Eddington regime has so far remained elusive. However, recent circumstantial evidence hinted at the presence of a recurrent evolutionary cycle in two archetypal ULXs: \ho\ and \ngc.}
   {We aim to build on our previous work and exploit the long-term high-cadence monitoring of \swift-XRT in order to provide robust evidence of the evolutionary cycle in these two sources and investigate the main physical parameters inducing their spectral transitions.}
   {We studied the long-term evolution of both sources using hardness-intensity diagrams (HID) and by means of Lomb-Scargle periodograms and Gaussian process modelling to look for periodic variability. We also applied a physically motivated model to the combined \chandra, \xmm, \nustar,\ and \swift-XRT data of each of the source spectral states.}
   {We robustly show that both sources follow a clear and recurrent evolutionary pattern in the HID that can be characterised by the hard ultra-luminous (HUL) and soft ultra-luminous (SUL) spectral regimes, and a third state with characteristics similar to the super-soft ultra-luminous (SSUL) state. The transitions between the soft states seem consistent with aperiodic variability, as revealed by a timing analysis of the light curve of \ho; albeit, further investigation is warranted. The light curve of \ngc\ shows a stable periodicity on a longer baseline of $\sim$ 200 days, possibly associated with the duration of the evolutionary cycle.}
   {The similarities between both sources provide strong evidence of both systems hosting the same type of accretor and/or accretion flow geometry. We support a scenario in which the spectral changes from HUL to SUL are due to a periodic increase of the mass-transfer rate and subsequent narrowing of the opening angle of the super-critical funnel. The narrower funnel, combined with stochastic variability imprinted by the wind, might explain the rapid and aperiodic variability responsible for the SUL--SSUL spectral changes. The nature of the longer periodicity of \ngc\ remains unclear, and robust determination of the orbital period of these sources could shed light on the nature of the periodic modulation found. Based on the similarities between the two sources, a long periodicity should be detectable in \ho\ with future monitoring.}

   \maketitle
%

\section{Introduction} \label{sec:intro}
Extensive monitoring of black hole (BH) and neutron star (NS) X-ray binaries has marked a breakthrough in our understanding of the accretion flow changes driving the flux variability in the sub-Eddington regime \citep[e.g.][]{done_observing_2003, remillard_x-ray_2006, gladstone_analysing_2007} thanks to the discovery of the hysteresis cycles. For instance, the discovery of the hysteresis cycle or `q-shape' diagram found in BH X-ray Binaries (BHXBs) \citep[e.g.][]{fender_towards_2004, belloni_states_2010} has allowed us to build a coherent picture of the physical changes in the accretion flow responsible for the multi-wavelength observational properties of each spectral state (associated with the presence or absence of radio jet emission, for instance). The importance of such evolutionary tracks is such that it is still used as a tool to identify accreting stellar-mass BHs today \citep[e.g.][]{zhang_nicer_2020}. For the same reason, the spectral changes observed in HLX-1 \citep{farrell_intermediate-mass_2009, godet_first_2009, servillat_x-ray_2011}, reminiscent of those observed in BHXBs but at three orders of magnitude higher in luminosity, remain one of the most compelling arguments in favour of the system harbouring an accreting intermediate-mass BH \citep[IMBH: $\sim$ 100 -- 10$^5$ M$_\odot$; see][for a review]{mezcua_observational_2017}.

At the same time, it was the atypical spectral states of ultra-luminous X-ray sources (ULXs), most of them being extragalactic, off-nuclear, point-like X-ray sources with an X-ray luminosity in excess of $\sim$ 10$^{39}$ erg/s \citep[see][for a review]{kaaret_ultraluminous_2017}, which suggested that the majority of these objects with L$_\text{X}$ $<$ 10$^{41}$ erg/s could be powered by super-Eddington accretion onto stellar-mass compact objects \citep[e.g.][]{stobbart_xmmnewton_2006, gladstone_ultraluminous_2009, sutton_ultraluminous_2013}. In this still poorly understood extreme regime, the intense radiation pressure caused by the high-mass transfer rate is expected to drive powerful and conical outflows \citep{shakura_black_1973,poutanen_supercritically_2007, ohsuga_why_2007, narayan_spectra_2017}. The atypical ULX spectral states were thus empirically classified as hard or soft ultra-luminous (HUL and SUL regimes), which were argued to depend on whether the observer had a clean view of the inner regions of the accretion flow or instead the source was viewed through the wind \citep{sutton_ultraluminous_2013}. The wind cone is expected to narrow with the accretion rate \citep[][]{king_masses_2009, kawashima_comptonized_2012} and may explain the spectral transitions frequently observed in ULXs \citep{sutton_ultraluminous_2013, middleton_spectral-timing_2015, gurpide_long-term_2021}. It has been argued that at higher inclinations or mass-accretion rates a ULX may appear as an ultra-luminous super-soft source \citep[ULSs; e.g.][]{urquhart_optically_2016}. This scenario is supported by the transitions from the SUL to the super-soft ultra-luminous (SSUL) regime observed in a handful of sources such as NGC 247 ULX-1 \citep[e.g.][]{feng_nature_2016, pinto_xmm-newton_2021} and NGC 55 ULX \citep{pinto_ultraluminous_2017}.

Multi-wavelength observations provide indirect evidence of such outflows or winds, in the form of inflated bubbles of ionised gas around a handful of ULXs \citep[e.g.][]{pakull_optical_2002, abolmasov_optical_2007} or optical \citep{fabrika_supercritical_2015} and X-ray spectral lines \citep[e.g.][]{pinto_thermal_2020}. The detection of X-ray pulsations in the light curve of six ULXs \citep{bachetti_ultraluminous_2013, furst_discovery_2016, israel_accreting_2017, carpano_discovery_2018, sathyaprakash_discovery_2019, rodriguez-castillo_discovery_2020} and possibly in another one \citep{quintin_new_2021} identified the accretor as an NS and provided direct evidence of super-Eddington accretion for at least a fraction of the objects. The existing evidence, together with the sustained luminosities above 10$^{39}$ erg/s over timescales of years \citep[e.g.][]{kaaret_x-ray_2009, grise_long-term_2013} and their relative proximity, make ULXs excellent laboratories for studying super-Eddington accretion. 

However, as opposed to Galactic X-ray binaries, the irregular monitoring of ULXs has precluded the study of any such evolutionary tracks, which are so important for our understanding of the accretion flow properties in the super-Eddington regime. For this reason, previous studies focussed mainly on discussing the different spectral states observed in ULXs \citep{sutton_ultraluminous_2013, middleton_spectral-timing_2015}. Therefore, these studies lacked information about the timescales and duration of each spectral state, that still hampers a full understanding of the nature of these transitions. 

For this reason, long-term monitoring programmes have proven particularly useful in our understanding of the mechanisms responsible for the variability in ULXs. These have revealed the presence of super-orbital periods on timescales of months (see e.g. \citet{vasilopoulos_ngc_2019} and \citet{townsend_orbital_2020} and references within), which are often variable \citep{kong_possible_2016, an_temporal_2016,vasilopoulos_m51_2020}. The nature of these super-orbital periods remains unclear, but it has recently been proposed that a Lense-Thirring precession of a super-critical disc and outflows \citep{middleton_lense-thirring_2018, middleton_lensethirring_2019} could explain the modulation seen in the light curves of PULXs \citep{dauser_modelling_2017}. This model has been proposed to explain the extreme flux modulation of factor $>$50 due to obscuration seen in the case of NGC 300\,ULX1, where a stable spin-up rate has been maintained during epochs of variable flux \citep{vasilopoulos_ngc_2019}. In addition, NGC 300\,ULX1 is perhaps the only ULX system where an orbital period has been proposed to be longer than one year based on both X-ray \citep{vasilopoulos_ngc_2018,ray_anti-glitches_2019} and infrared properties \citep{heida_discovery_2019, lau_uncovering_2019}.

In \citet{gurpide_long-term_2021}, we studied the long-term variability of a sample of 17 ULXs in order to understand the accretion flow geometry that could be responsible for the spectral variability observed in each source. In particular, we showed that certain ULXs followed an interesting evolutionary track in the hardness-luminosity diagram (HLD). Of particular interest were two archetypal ULXs, \ngc\ and \ho, whose evolution could be clearly summarised in three states: `hard-intermediate', `soft-bright' and `soft-dim' \citep[see Figure 6 from][]{gurpide_long-term_2021}. 

However, the limited cadence offered by \xmm\ and \chandra\ precluded us from constraining the exact temporal evolution of the sources and therefore the nature of these transitions. In this work, we build on that  of our previous paper by exploiting the high-cadence monitoring offered by \swift-XRT \citep{burrows_swift_2005} in order to constrain the temporal timescale of the transitions. We will show that these two sources describe a very similar and well-defined recurrent evolutionary cycle. Moreover, we show that the whole cycle is likely periodic, providing a clear connection between long-term periodicities and spectral changes in these ULXs.

Section \ref{sec:data_reduction} outlines the data reduction process. Section \ref{sec:data_analysis} presents the data analysis and results. In Section \ref{sec:discussion}, we discuss our findings. Finally, in Section \ref{sec:conclusions} we summarise our conclusions.
\section{Data reduction}\label{sec:data_reduction}
For this work, we used the dataset analysed in \citet{gurpide_long-term_2021} to which we added long-term monitoring provided by archival \swift-XRT data. The details of our data reduction for \xmm\ \citep{jansen_xmm-newton_2001}, \nustar\ \citep{harrison_nuclear_2013}, and \chandra\ \citep{weisskopf_chandra_2000} can be found in \citet{gurpide_long-term_2021}. Here, we briefly summarise the main steps followed. The data used in this work is summarised in Table \ref{tab:states}.

\begin{table*}  
 \centering 
 \caption{Observation log used for each state.}\label{tab:states}
 \begin{tabular}{lcccccc} 
 \hline\hline
  \noalign{\smallskip} 
 State & Telescope &Obs. ID & Date  & Good Exp &  Detector/Filter\tablefootmark{a}& ($M$--$S$)/$T$\tablefootmark{b} \\ 
  &  & &  & (ks) & &  \\ 
 \hline \hline 
 \noalign{\smallskip}
\multicolumn{7}{c}{\ngc} \\ 
\hline
 \noalign{\smallskip} 
\multirow{5}{*}{SUL} &\xmm\ &0405690101 & 2006-11-16 & 7.82/5.85/5.82 & pn/MOS1/MOS2 & --0.32
\\  
& \chandra & 3936 & 2003-08-14 & 4.56 & ACIS & --0.49
\\ 
& \chandra & 3937 & 2003-08-17 & 4.64 & ACIS & --0.49
\\ 
& \chandra& 3941 & 2003-09-14 & 4.88 & ACIS & --0.43
\\ 
 & \chandra & 3943 & 2003-10-03 & 4.90 & ACIS &--0.44

 \\ 
 \noalign{\smallskip}
\hline
 \noalign{\smallskip}
\multirow{5}{*}{SSUL} & \chandra & 2029 & 2001-05-02 & 9.01 & ACIS & --0.62
 \\
& \chandra & 3934 & 2003-08-09 & 4.74 & ACIS & --0.73
 \\ 
& \chandra & 3935 & 2003-08-11 & 4.5 & ACIS  & --0.61
\\ 
& \chandra & 3938 & 2003-08-19 & 5.15 & ACIS & --0.73
 \\ 
& \chandra & 3939 & 2003-08-27 & 5.14 & ACIS & --0.79
 \\ 
 \noalign{\smallskip}
\hline
\noalign{\smallskip}
\multirow{6}{*}{HUL} & \xmm\ &0142770101 & 2003-01-06 & 15.33/18.49/18.50 &  pn/MOS1/MOS2 & -0.26
\\ 
&\xmm\ & 0741960101 & 2014-06-27 & 19.01/22.54/22.73 & pn/MOS1/MOS2 & --0.29
\\  
&\xmm\ & 0693851401 & 2013-04-21 & 13.37/16.40/16.46 & pn/MOS1/MOS2 & --0.28\\  
&\xmm\ & 0693850701 & 2013-04-29 & 9.97/14.24/15.45 & pn/MOS1/MOS2 & --0.29

\\  
&  \nustar& 30002037002 & 2013-04-19 & 95.96/95.80 & FPMA/FPMB &\\ 
&  \nustar & 30002037004 & 2013-04-29 & 88.98/88.85 & FPMA/FPMB &\\ 
 \noalign{\smallskip} 
 \hline 
 \hline 
  \noalign{\smallskip} 
 \multicolumn{7}{c}{\ho} \\
 \hline 
 \noalign{\smallskip} 
\multirow{3}{*}{SUL}& \xmm\ & 0200470101 & 2004-04-15 & 34.95/49.61/51.10 & pn/MOS1/MOS2 & --0.27
\\ 
& \xmm\ & 0112520601 & 2002-04-10 & 4.64/9.45/9.65 & pn/MOS1/MOS2 & --0.31\\ 
&\swift& 00035475019 & 2009-12-01 & 23.40 & XRT & \\ 
  \noalign{\smallskip} 
 \hline  \noalign{\smallskip} 
\multirow{1}{*}{SSUL} & \xmm\ & 0112520901 & 2002-09-18 & 3.78/5.85/6.45 & pn/MOS1/MOS2 & -0.51
 \\ 
  \noalign{\smallskip} 
 \hline  \noalign{\smallskip} 
\multirow{9}{*}{HUL}  & \xmm\ & 0724810101 & 2013-09-09 & 4.65/6.63/6.63 & pn/MOS1/MOS2 & --0.25
 \\ 
& \xmm\ & 0724810301 & 2013-09-17 & 5.81/8.61/9.31 & pn/MOS1/MOS2 & --0.29
 \\ 
& \nustar & 30001031002 & 2013-09-09 & 31.38/31.32 & FPMA/FPMB &\\ 
& \nustar & 30001031003 & 2013-09-09 & 79.44/79.34 & FPMA/FPMB &\\ 
& \nustar &30001031005 & 2013-09-17 & 111.10/111.00 & FPMA/FPMB & \\
& \textit{HST}& j9dr03r1q & 2013-08-24 & 1.505 & ACS/WFC/F550M &\\
& \textit{HST}& icdm60ipq & 2006-01-28 & 2.424 & WFC3/UVIS/F275W &\\
& \textit{HST}& icdm60j0q & 2013-08-24 & 1.146 & WFC3/UVIS/F336W &\\
& \textit{HST}& icdm60j6q & 2013-08-24 & 0.992 & WFC3/UVIS/F438W & \\
 \noalign{\smallskip} 
 \hline 
 \hline 
 \end{tabular} 
 \tablefoot{\tablefoottext{a}{Only for the \hst\ data.} \tablefoottext{b}{Softness computed as per \cite{urquhart_optically_2016}, but only for \xmm\ and \chandra\ observations. $M$, $S,$ and $T$ refer to the net count rates in the 0.3--1.1 keV, 1.1--2.5 keV, and 0.3--10 keV, respectively.}}
 \end{table*}

\subsection{\xmm}
For the \epicpn\ \citep{struder_european_2001} and MOS \citep{turner_european_2001} cameras we selected events with the standard filters, removing pixels flagged as bad and those close to a CCD gap. We filtered periods of high background flaring by removing times where the count rate was above a chosen threshold by visually inspecting the background high-energy light curves. Source products were extracted from a circular region with a radius of 40" and 30" for pn and MOS, respectively. The background region was selected from a larger, circular, source-free region on the same chip as the source (when possible). We generated response files using the tasks \texttt{rmfgen} (version 2.8.1) and  \texttt{arfgen} (version 1.98.3). Finally, we regrouped our spectra to have a minimum of 20 counts per bin to allow the use of \chisq statistics and also avoid oversampling the instrumental resolution by setting a minimum channel width of 1/3 of the FWHM energy resolution.

\begin{table} 
 \centering 
 \caption{\label{tab:observations_chandra}\chandra\ observations of \ngc\ considered in this work that were not included in \cite{gurpide_long-term_2021}. The number of Cs indicates the number of observations fitted together after checking their consistency.} 
  
\begin{tabular}{ccccc} 
 \hline
 \noalign{\smallskip}
 Epoch &Date & Obs. ID  & Good Exp. & Total Counts \\ 
 &  &  & (ks) &\\  
 \hline
 \noalign{\smallskip}
\multirow{2}{*}{1CC} & 2003-08-09 &3934 &4.7& 539 \\
&2003-08-11 &3935 &4.5 &616  \\  
2C & 2003-08-19 & 3938& 5.1 & 567  \\ 
3C & 2003-08-27 & 3939&5.1 & 586\\
\hline
 \end{tabular} 
 \end{table}
\subsection{\chandra}
All \chandra\ data were reduced using the \textsc{CIAO} software version 4.11 with calibration files from \textit{CALDB} 4.8.2. We used extraction regions given by the tool \textit{wavdetect} to extract source events. Background regions were selected from circular nearby source-free regions that were roughly three times larger. For this work, we included four \chandra\ observations of \ngc\, that were not considered by \cite{gurpide_long-term_2021} due to their limited number of counts ($<$ 700). These are presented in Table \ref{tab:observations_chandra}. All data were re-binned to a minimum of 20 counts per bin so as to use \chisq\ statistics, except for these low count observations that are re-binned to two counts per bin and analysed using the pseudo $C$-statistic \citep{cash_parameter_1979}, $W$-stat in \xspec\ \footnote{\url{https://heasarc.gsfc.nasa.gov/xanadu/xspec/manual/XSappendixStatistics.html}}, which is suitable for the analysis of low count spectra when the background is not being modelled. The choice of two counts per bin was made in order to reduce the biases associated with the $W$-stat\footnote{ \url{https://giacomov.github.io/Bias-in-profile-poisson\\-likelihood/}}. We note that $W$-stat is the default statistic used when the $C$-stat option is selected and a background file is read in.
\subsection{\nustar}
The \nustar\ data were extracted with the \nustar\ Data Analysis Software version 1.8.0 with \textit{CALDB} version 1.0.2. Source and background spectra were extracted using the task \textit{nuproducts} with the standard filters. Source events were selected from a circular of $\sim$ 60" centred on the source. Background regions were selected from larger, circular, source-free regions and on the same chip as the source but as far away as possible to avoid contamination from the source itself. We re-bin \nustar\ data to 40 counts per bin, owing to the lower energy resolution compared to the \textit{EPIC} cameras.

\subsection{\swift}
 We retrieved all publicly available snapshots from the UK \swift-XRT data centre\footnote{\url{https://www.swift.ac.uk/user_objects/}} as of 1 August 2020. Each \swift\ observation is split into several snapshots, each of which we considered separately in our periodicity analysis (Section \ref{sub:periodicity_analysis}). In order to build products (i.e. light curves and/or spectra), we used the standard tools outlined in \citet{evans_online_2007, evans_methods_2009}, considering all detections above the 95\% confidence level. We further discarded two snapshots that had a signal-to-noise below 2 from the dataset of \ngc\ for our periodicity analysis (Section \ref{sub:periodicity_analysis}). We also extracted the count rate in two energy bands, a soft band (0.3 -- 1.5 keV range) and a hard band (1.5 -- 10 keV range), to build a hardness--intensity diagram (HID) (Section \ref{sec:long_term_variability}).  We were left with 152 snapshots with a mean cadence between observations of 4.7 days for \ngc\ and 303 snapshots with a mean cadence between observations of 1.6 days for \ho.
\section{Data analysis and results}\label{sec:data_analysis}
\subsection{Long-term variability}\label{sec:long_term_variability}
\begin{figure*}
    \centering
    \includegraphics[width=\textwidth, height=0.45\textheight]{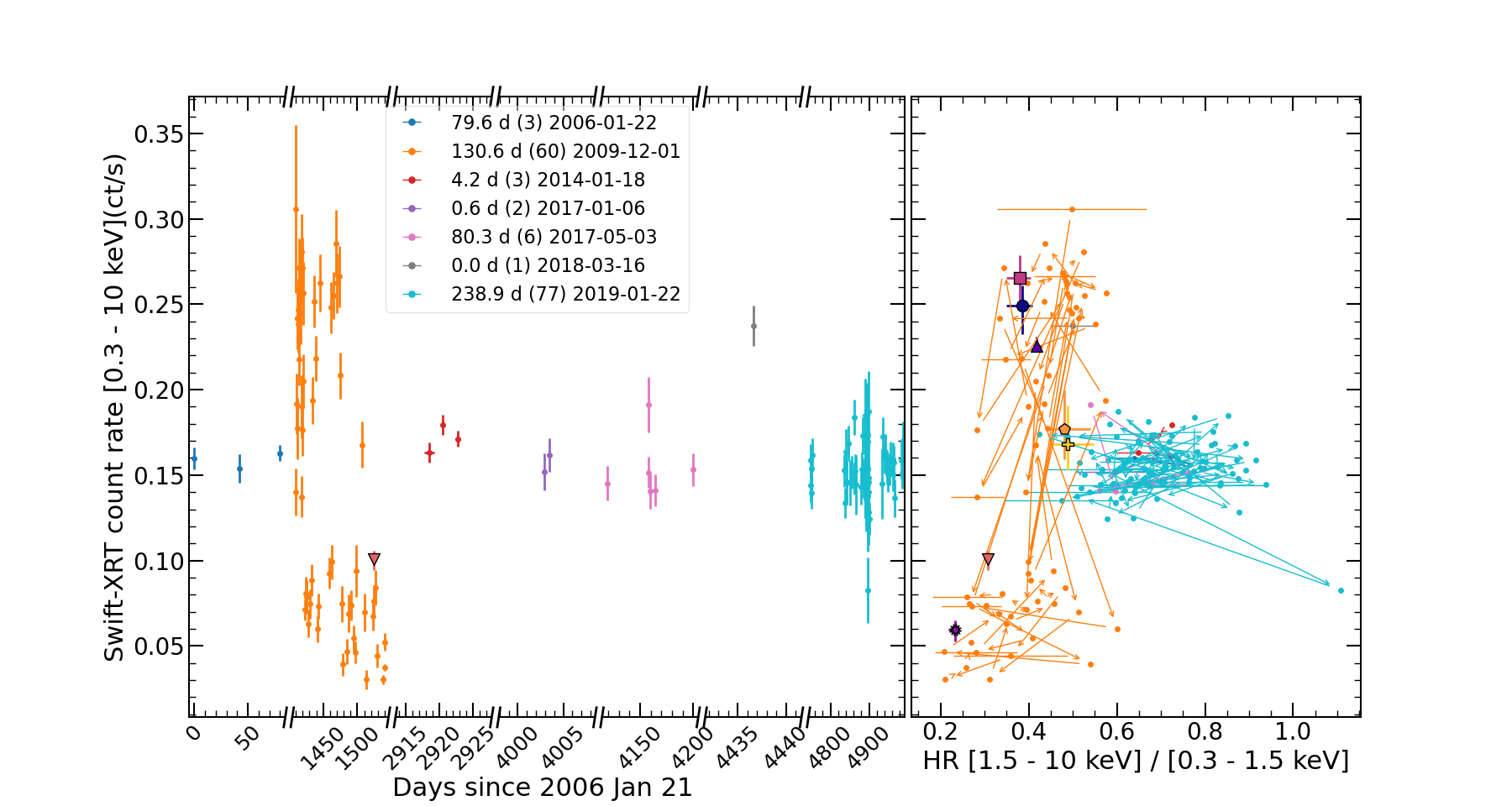}
    \includegraphics[width=\textwidth, height=0.45\textheight]{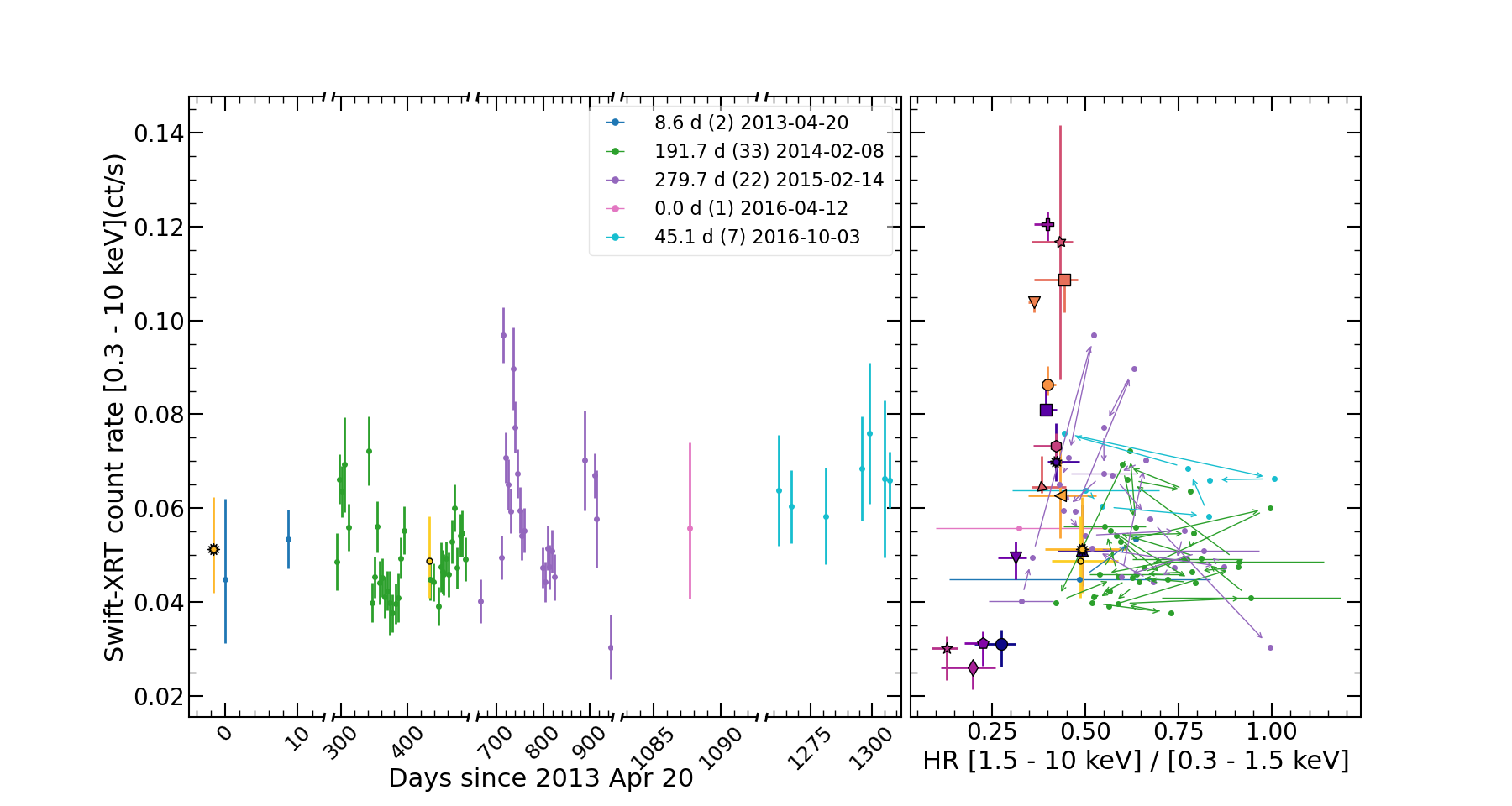}
    \caption{\textit{Left}: \swift-XRT light curve in the 0.3 -- 10 keV band for \ho\ (top) and \ngc\ (bottom) with its associated HID (right). Continuous monitoring intervals with data gaps of less than 12 weeks have been coloured accordingly and the legend indicates the duration in days, the number of observations, and starting date of each continuous period. Arrows in the HID indicate chronological order and errors are given every eight datapoints for clarity. Big coloured markers correspond to the \swift-XRT 0.3--10 keV count-rate estimates of the \xmm\ and \chandra\ observations based on best-fit models from \cite{gurpide_long-term_2021} (see text for details). Errors are at 1 $\sigma$ confidence level. We note that the X-axis has been split on the left panels. }
    \label{fig:swift_data}
\end{figure*}

In order to investigate the long-term variability of both sources, we inspected the \swift-XRT light curves and built HIDs for both sources (Figure \ref{fig:swift_data}). Here, we considered the data binned by observations to ensure that the source was well detected in each of the energy bands. 

The \swift-XRT data for \ho\ reveal two distinct trends, one from the 2009--2010 data in which the source switched quickly and repeatedly, within a few days \citep[see also][]{grise_x-ray_2010}, between a bright and soft state (soft-bright state: HR $\sim$ 0.5, count rate in the 0.3 -- 10 keV band $\sim$ 0.28 cts/s) and a dim and softer state (soft-dim state: HR $\sim$ 0.25, count rate of $\sim$ 0.05 cts/s). This variability is also observed intra-observation in the long \citep[30 ks good exposure;][]{gurpide_long-term_2021} obs id 0561580401 (this observation is indicated as a pink triangle pointing downwards in Figure \ref{fig:swift_data}). 

\begin{figure}
    \centering
    \includegraphics[width=0.48\textwidth]{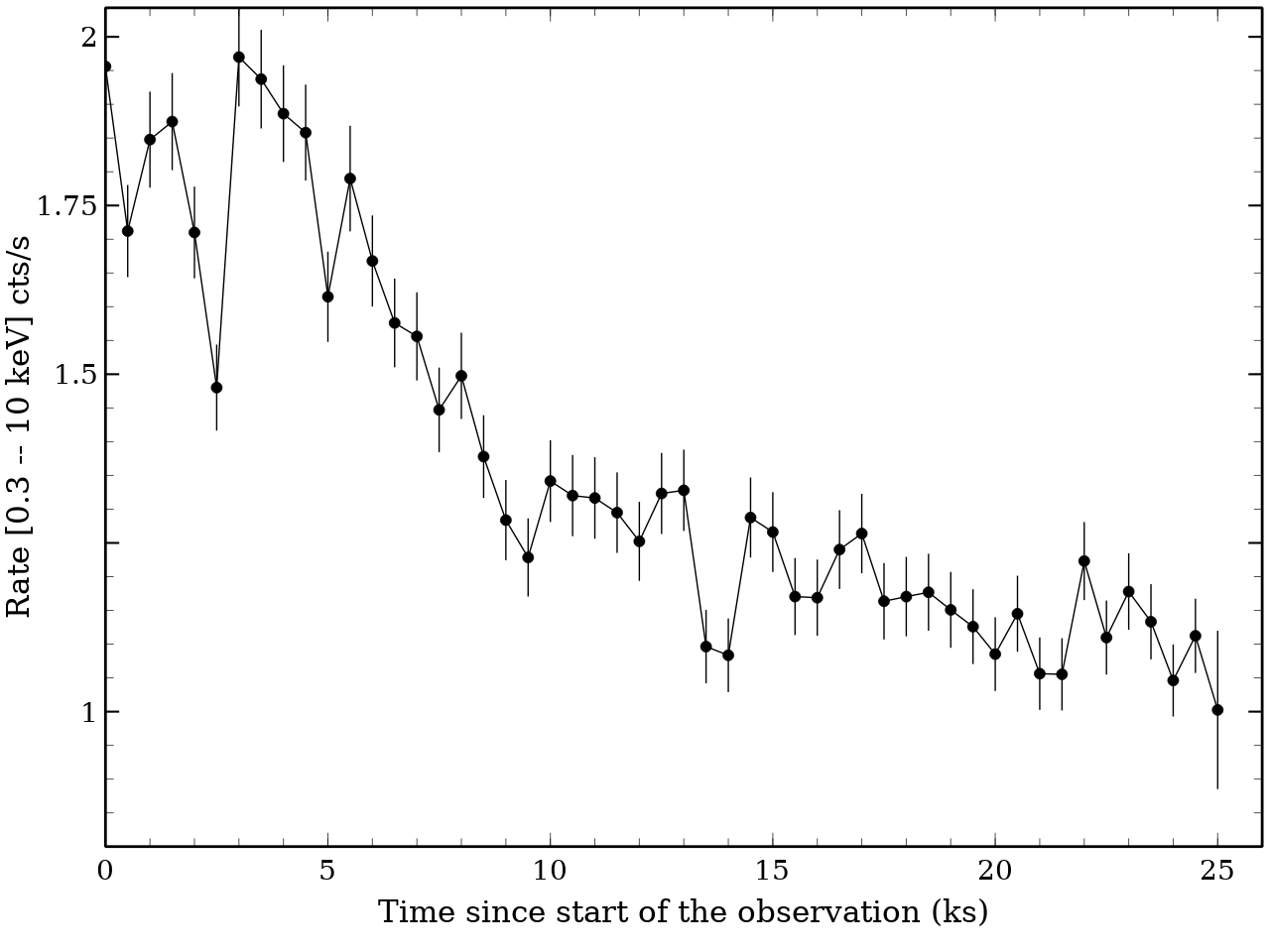}
    \caption{0.3 -- 10 keV background-subtracted EPIC-pn light curve of \ho\ (obs id 0561580401) binned to 500 s and corrected using the task \texttt{epiclccorr}.}
    \label{fig:light curve}
\end{figure}

The second trend is well sampled in the 2019 monitoring and shows that the source remained stable with HR $\sim$ 0.75 and a 0.3--10 keV count rate of $\sim$ 0.15 -- 0.2 cts/s (hard-intermediate state). The source has been repeatedly observed in this state on several occasions prior to 2019, in 2006, 2014, and 2017. From the 2009--2010 monitoring, the duration of whole soft state transitions seems to last at least 131 days, whereas the 2019 monitoring indicates that the duration of the hard state seems to be of at least 240 days.

\begin{figure}
    \centering
    \includegraphics[width=0.49\textwidth]{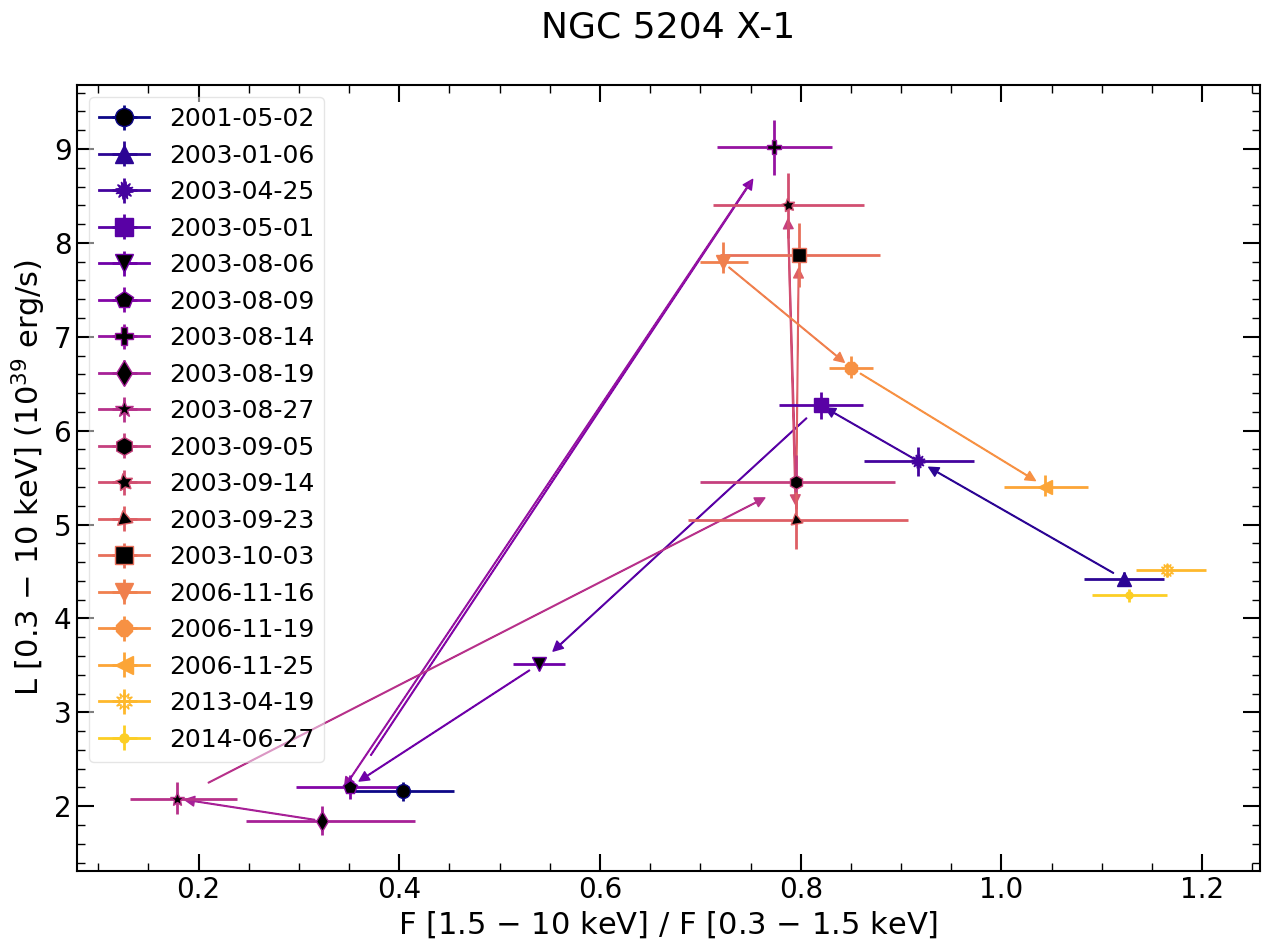}
    \caption{Temporal tracks on the HLD adapted from \cite{gurpide_long-term_2021} for \ngc. New \chandra\ observations have been included to illustrate the short-term variability in 2003 (see Tables \ref{tab:observations_chandra} and \ref{tab:fit_chandra}). Black and coloured markers indicate \chandra\ and \xmm\ observations, respectively. Arrows indicate the chronological order and are only plotted for observations closer in time than four months.}
    \label{fig:hld_ngc5204x1}
\end{figure}

The identification of any pattern in the \swift-XRT data is not straightforward in the case of \ngc\ because \swift\ caught the source mostly in the hard-intermediate state, albeit \cite{gurpide_long-term_2021} showed that \ho\ and \ngc\ evolved in a similar manner from the analysis of archival \xmm\ and \chandra\ data. To illustrate this, we adapted Figure 6 from \cite{gurpide_long-term_2021} in our Figure \ref{fig:hld_ngc5204x1} by considering the full \chandra\ monitoring of \ngc\ presented by \cite{roberts_chandra_2006} (see Table \ref{tab:observations_chandra}) to highlight the short-term variability the source underwent in 2003. We fitted these new \chandra\ observations using the same model as in \cite{gurpide_long-term_2021}, an absorbed dual-thermal component in \xspec\ \citep{arnaud_xspec:_1996} version 12.10.1f. We fixed the galactic \nh\ and the local \nh\ to the same values found in \cite{gurpide_long-term_2021}. The best-fit parameters are summarised in Table \ref{tab:fit_chandra}. For our aim, we retrieved unabsorbed fluxes using \cflux\ in the 0.3 -- 1.5 keV and the 1.5 -- 10 keV bands, and we show them together with the previous observations from \cite{gurpide_long-term_2021} in Figure \ref{fig:hld_ngc5204x1}. 

To put the \xmm\ and \chandra\ observations from \cite{gurpide_long-term_2021} in perspective with respect to the \swift-XRT data, we estimated the count rates in the 0.3 -- 1.5 keV, 1.5 -- 10 keV and 0.3 -- 10 keV bands from the best-fit model parameters from \cite{gurpide_long-term_2021} by convolving each model with the latest redistribution matrix (swxpc0to12s6\_20130101v014.rmf) and ancillary file (swxs6\_20010101v001.arf) available as of 1 August 2020. In order to derive the uncertainties, we linearly interpolated the 1$\sigma$ uncertainties from the observed fluxes to convert them to equivalent \swift-XRT count rates. This is represented in Figure \ref{fig:swift_data} using the same colour code as in \cite{gurpide_long-term_2021} in the case of Holmberg II X-1 and that of Figure \ref{fig:hld_ngc5204x1} in the case of \ngc. 

\begin{table*} 
 \centering 
 \caption{Results of the fit of the \chandra\ observations considered in this work that were not included in \cite{gurpide_long-term_2021}. Errors are quoted at the 90\% confidence level. The number of Cs indicates the number of observations jointly fitted.} \label{tab:fit_chandra}
\begin{tabular}{lcccccccc} 
 \hline
 \noalign{\smallskip}
 Epoch & $kT_\text{soft}$ & norm & $kT_\text{hard}$ & norm  & L$_\text{soft}^a$& L$_\text{hard}^a$ & L$_\text{tot}^a$ & $C$-stat(dof) \\ 
&  keV &  & keV & 10$^{-2}$ &  (0.3--1.5 keV)  & (1.5--10 keV) & (0.3 -- 10 keV)  &  \\ 
 \hline
 \noalign{\smallskip}
 \multicolumn{9}{c}{NGC 5204 X$-$1}\\
1CC & 0.26$\pm$0.03& 8\errors{6}{3}& 1.0\errors{0.5}{0.2}& 2\errors{4}{2}& 1.6$\pm$0.1& 0.57\errors{0.09}{0.08}& 2.2$\pm$0.1 &1.24/218\\ 
 2C &0.28$\pm$0.04& 6\errors{6}{3}& 1.2\errors{b}{0.5}& 0.50\errors{1.7}{0.5}& 1.4$\pm$0.1& 0.4$\pm$0.1& 1.8$\pm$0.2 & 1.03/104\\ 
 3C & 0.26$\pm$0.04& 10\errors{9}{4}& 0.9\errors{b}{0.4}& 1\errors{26}{1}&1.8\errors{0.2}{0.1}& 0.32\errors{0.10}{0.08}& 2.1$\pm$0.2 &0.79/94 \\
  \noalign{\smallskip}
\hline
 \end{tabular} 
 \tablefoot{The Galactic \nh\ and the local \nh\ were frozen to 1.75$\times$10$^{20}$ cm$^{-2}$ \cite{kalberla_leiden/argentine/bonn_2005} and to 3.0$\times$10$^{20}$ cm$^{-2}$, respectively, following \cite{gurpide_long-term_2021}. \tablefoottext{a} In units of 10$^{39}$ erg/s. \tablefoottext{b}     We set an upper limit of 5 keV.} 
 \end{table*}

The pattern revealed by the \chandra\ monitoring is reminiscent of the transitions observed in the 2009--2010 \swift-XRT data of \ho\ (Figures \ref{fig:swift_data} and \ref{fig:hld_ngc5204x1}). Once again, the source transits from soft-bright (HR $\sim$ 0.8, unabsorbed L$_\text{X}$ $\sim$ 9 $\times$ 10$^{39}$ erg/s) to soft-dim (HR $\sim$ 0.2, unabsorbed L$_\text{X}$ $\sim$ 2 $\times$ 10$^{39}$ erg/s) on several occasions over $\sim$ 3 months. The tightest constraints on the duration of these transitions come from the 2003-08-09/19 observations, which show that the source went back and forth from the soft-dim state to the soft-bright state in ten days. The \xmm\ observations from Figure \ref{fig:hld_ngc5204x1} (coloured markers) show how the source went from the hard-intermediate state (epoch 2003-01-06) to the soft-bright state and started transiting from soft-bright to soft-dim from epoch 2003-05-01 onwards. Another transition from bright/soft to hard-intermediate was also observed by \xmm\ in 2006. Given that the source was found in the soft-dim state in 2001 and back in that state in 2003-08-06 (Figure \ref{fig:hld_ngc5204x1}), the full cycle seems to have a duration of less than 2 years. It is also observed that \ngc\ was not detected in the soft-dim state by \swift-XRT.

Overall, the multi-instrument data presented reveal that the evolution of both \ngc\ and \ho\ are akin and can be characterised in three states: soft-bright, soft-dim, and hard-intermediate. The 2006 transitions of \ngc\ were also discussed by \citet{sutton_ultraluminous_2013}, and thus it is easy to see that the soft-bright state corresponds to the soft ultra-luminous (SUL) regime defined by these authors, whereas the hard-intermediate states correspond to the hard ultra-luminous (HUL) regime. The soft-dim state instead shows several characteristics similar to the SSUL regime \citep{feng_nature_2016, urquhart_optically_2016, pinto_ultraluminous_2017}, namely a luminosity around 10$^{39}$ erg/s (Figure \ref{fig:hld_ngc5204x1}) and little emission above 2 keV \citep[see Figure 6 in][]{gurpide_long-term_2021}. To illustrate this further, we measured net count rates in the same three energy bands as \cite{urquhart_optically_2016}: 0.3.--1.1 keV ($S$), 1.1--2.5 keV ($M$), and 0.3--7.0 keV ($T)$. We computed $(M - S)/T$ for each of the \xmm\ and \chandra\ observation available for each state (Table \ref{tab:states}). We see that the values for the soft-dim state of \ngc\ reach $(M - S)/T$ $\approx$ -- 0.8, the approximate threshold the authors used to classify ULX sources as super-soft. The lowest value of \ho\ registered by \xmm\ is --0.5, which overlaps with some super-soft sources, depending on the model considered (see their Figure 1). We also note that \swift\ observed \ho\ at even lower fluxes than those observed by \xmm\ (Figure \ref{fig:swift_data}) and that \ho\ is generally softer than \ngc \citep{gurpide_long-term_2021}, suggesting that these differences in $(M - S)/T$ might be mostly instrumental and that the properties of the SSUL state are broader than based on hardness alone. We discuss in Section \ref{sec:discussion} why \ho\ and \ngc\ appear slightly harder in the SSUL state than canonical ULS sources.

Additional evidence for the similarity between these states and the ULS regime comes from the rapid $<$ 2-day variability observed in the soft-bright to soft-dim transitions in the \swift-XRT light curve of \ho\ \citep[see also][]{grise_x-ray_2010}. This is reminiscent of the transitions observed in the ULS in M81 \citep{liu_no_2008}, which shows sudden flux drops and rises on timescales of $\sim$ 1 ks, similarly to those seen in \ho\ (Figure \ref{fig:light curve}). The ULX-ULS transitions seen in NGC 247 ULX-1 and NGC 55 ULX are also associated with the brightest states \citep{feng_nature_2016, pinto_ultraluminous_2017, pinto_xmm-newton_2021}, which is exactly what is observed in \ho\ and \ngc. Thus, \ho\ and \ngc\ may be the first ULXs observed to switch through all these three canonical ULX spectral states. Hereafter, we use the terms SUL, HUL, and SSUL to refer to the spectral states seen in \ho\ and \ngc\ for consistency with previous works.      

\subsection{Periodicity analysis} \label{sub:periodicity_analysis}
\subsubsection{Lomb-Scargle periodograms}
To investigate the presence of any periodicity in the long-term evolution of \ngc\ and \ho, we performed Lomb-Scargle periodograms\footnote{\url{https://docs.astropy.org/en/stable/timeseries/lombscargle.html}\citep{vanderplas_understanding_2018}} \citep{lomb_least-squares_1976,scargle_studies_1982}, which is suitable to look for stable sinusoidal periodicities. In this implementation of the Lomb-Scargle periodogram, the uncertainties in the measurements are used to weight the differences between the sinusoidal model and the data, as with a classical least-squares estimation. We also adopted the floating-mean periodogram, in which the mean count rate is fitted in the model instead of being subtracted as for the classical periodogram, which has been shown to give higher accuracy, especially when the data does not have full phase coverage \citep{vanderplas_understanding_2018}. For this analysis, we used the data from the \swift-XRT as it has the highest cadence and longest monitoring.

In the case of unevenly sampled data, there is no well-defined Nyquist frequency and thus the maximum frequency explored in the periodogram has to be set according to some physical and/or observing constraints. In this case, we restricted the maximum frequency explored to 5 $^{-1}$days, which is slightly above the typical cadence at which \swift\ visited each source. The minimum frequency explored is set to  1/T$_\text{segment}$ where T$_\text{segment}$ is the duration of the continuous light-curve segment considered in each case.  We oversampled the periodogram by a factor of five (i.e. $\Delta f$ = $1/nT_\text{segment}$ where n = 5).

\subsubsection{False-alarm probability: Bootstrap method}
A common challenge in the Lomb-Scargle periodogram is to estimate the significance of a peak in the power spectrum. In the presence of white noise,  \cite{scargle_studies_1982} showed that the false-alarm probability of a single peak with power $P$ is $F(P) = 1 - e^{-P}$. However, because we are essentially trying to detect a peak over a certain frequency range, one should account for the fact that several frequencies are being explored. This is challenging because the frequencies in the Lomb-Scargle periodogram are not independent, and thus one cannot compute the false-alarm probability simply by assuming that several independent frequencies (i.e. trials) have been looked for. Some empirical estimates of the number of independent frequencies exist \citep[e.g.][]{horne_prescription_1986}, but the most robust method is to rely on Monte Carlo simulations to calibrate the reference distribution as it makes few assumptions on the number of independent frequencies and considers the observing cadence of the real data, so the effects of the observing window are taken into account.

We thus estimated 2$\sigma$ and 3$\sigma$ false alarm probabilities in the absence of a signal by using the bootstrap method outlined in \cite{vanderplas_understanding_2018}. This method generates N light curves by drawing random samples from the original measurements to generate light curves with the same time coordinates as for the original data. For each randomised light curve, the same Lomb-Scargle periodogram is performed as for the real data. From the maxima of these fake periodograms, the reference distribution is built. Another advantage of this method is that the null-hypothesis model is not based on pure white noise, since the reference distribution is built on randomised data, although there are other limitations to consider (which we discuss in Section \ref{sec:redfit}). We performed 3500 simulations to ensure that at least ten peaks are found above the 3$\sigma$ level and thus that the reference distribution was well sampled \citep{vanderplas_understanding_2018}.

\begin{figure*}
     \centering
     \includegraphics[width=0.49\textwidth]{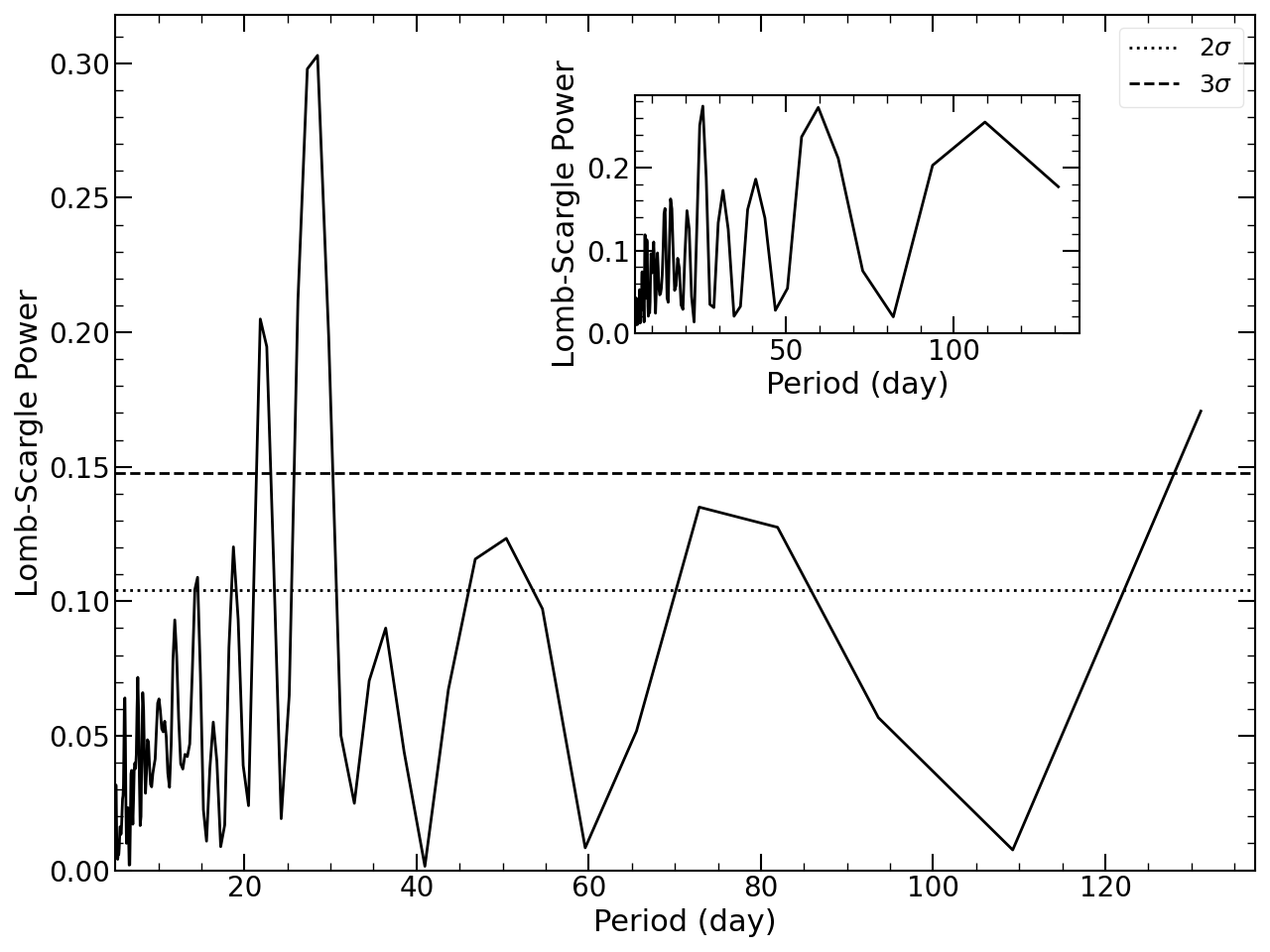}
     \includegraphics[width=0.49\textwidth]{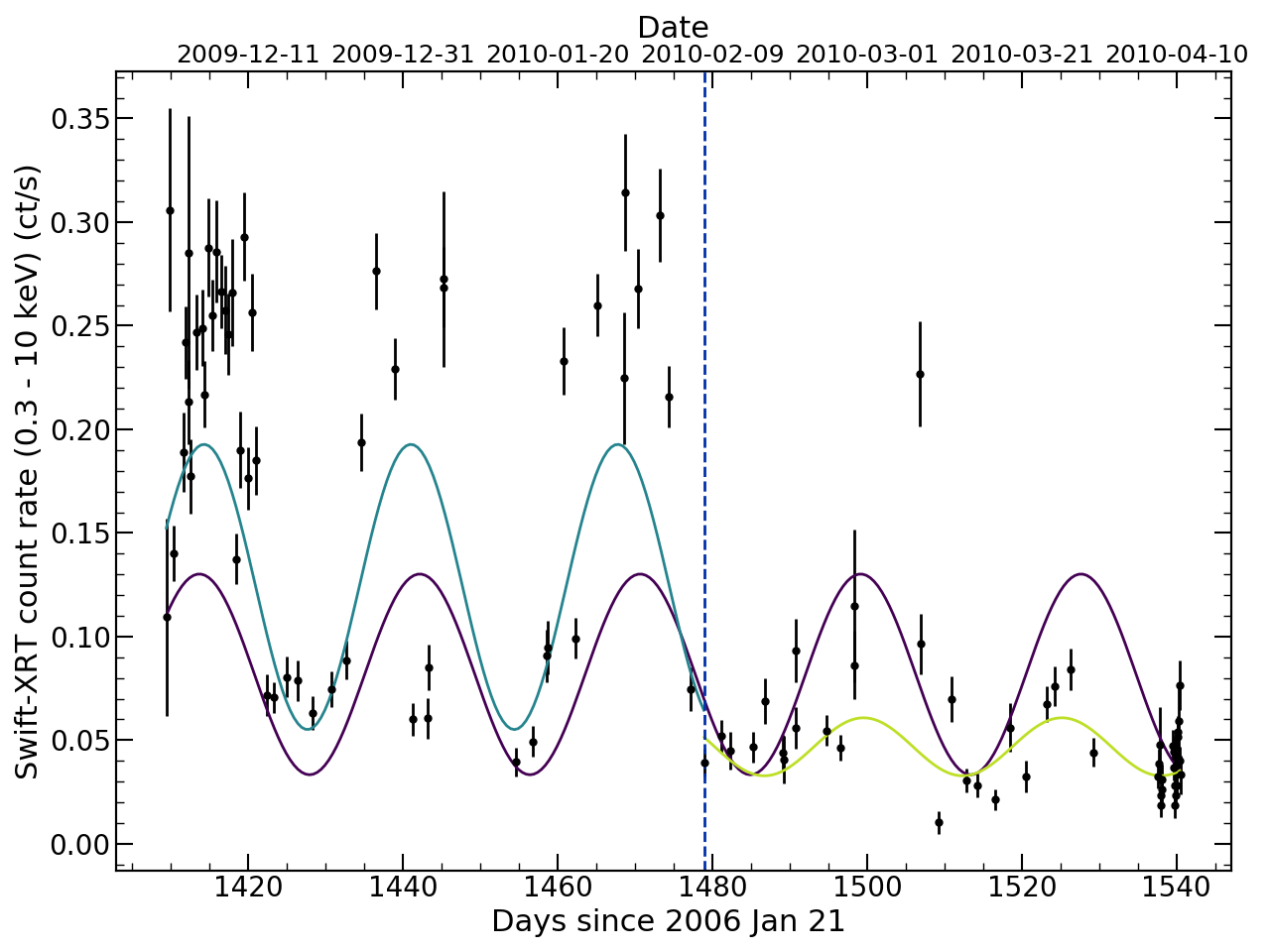}
    \includegraphics[width=0.49\textwidth]{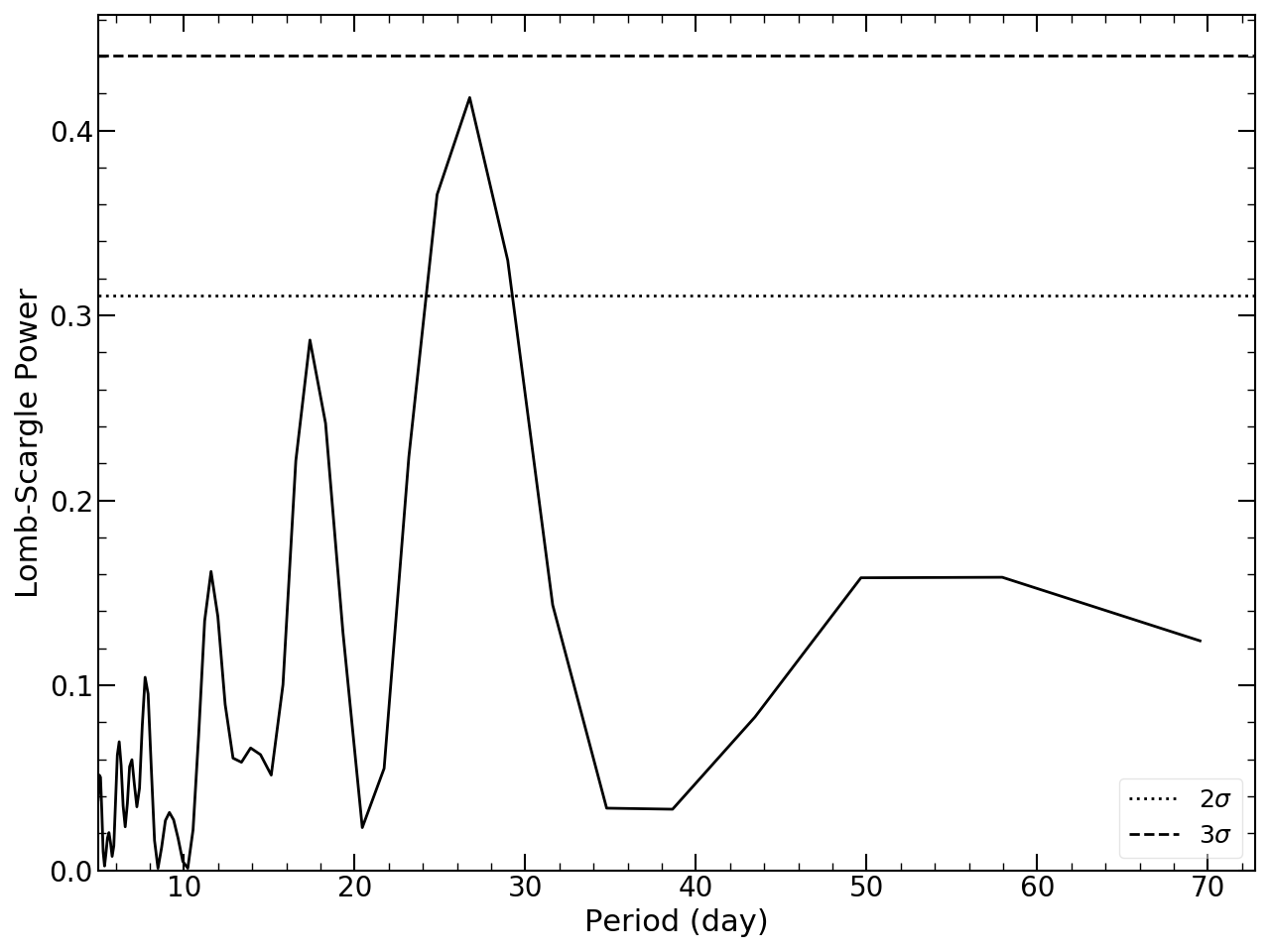}
    \includegraphics[width=0.49\textwidth]{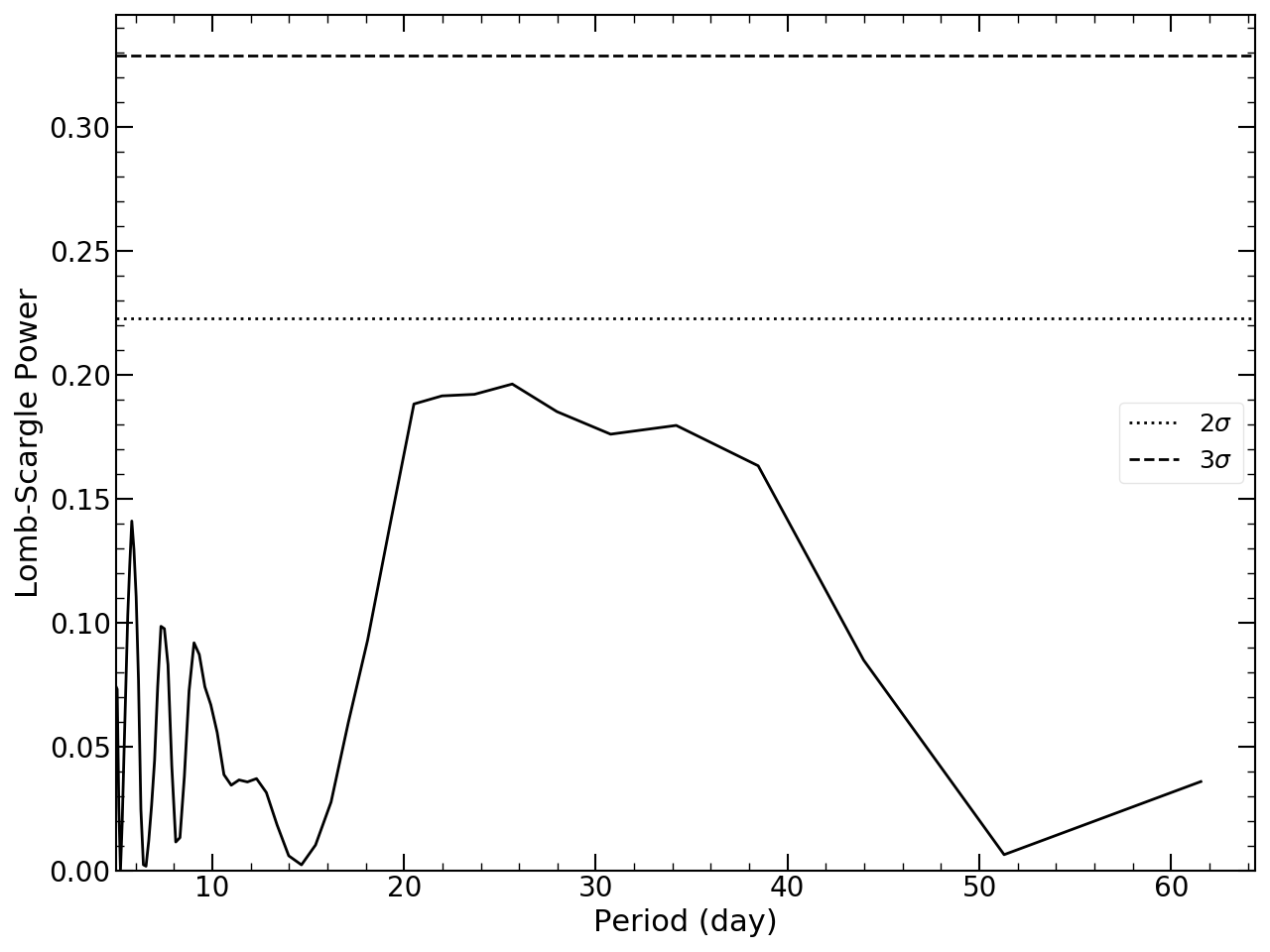}
     \caption{Top left: Lomb-Scargle periodogram for the entire 2009--2010 \swift-XRT dataset of Holmberg II X-1 (black solid line). The 2$\sigma$ and 3$\sigma$ levels (dashed and dotted lines, respectively) mark the false alarm probability computed from bootstrapped samples (see text for details). The inset shows the structure of the observing window. Top right: \swift-XRT light curve corresponding to the 2009--2010 segment. The solid purple line indicates the best-fit period from the periodogram of the whole light-curve segment. The solid light blue line indicates the best-fit period from the periodogram considering only the light curve segment to the left of the blue dashed line, while the yellow solid line indicates the best-fit period from the light-curve segment to the right of the blue dashed line. Bottom: Lomb-Scargle periodograms of two sub-segments of the 2009--2010 data separated by the blue dashed line in the top right panel (left and right panels correspond to the data prior to and after  $\sim$ day 1480).}
     \label{fig:hoIIx1_periodogram_2009_2010}
 \end{figure*}

\begin{figure}
     \centering
     \includegraphics[width=0.49\textwidth]{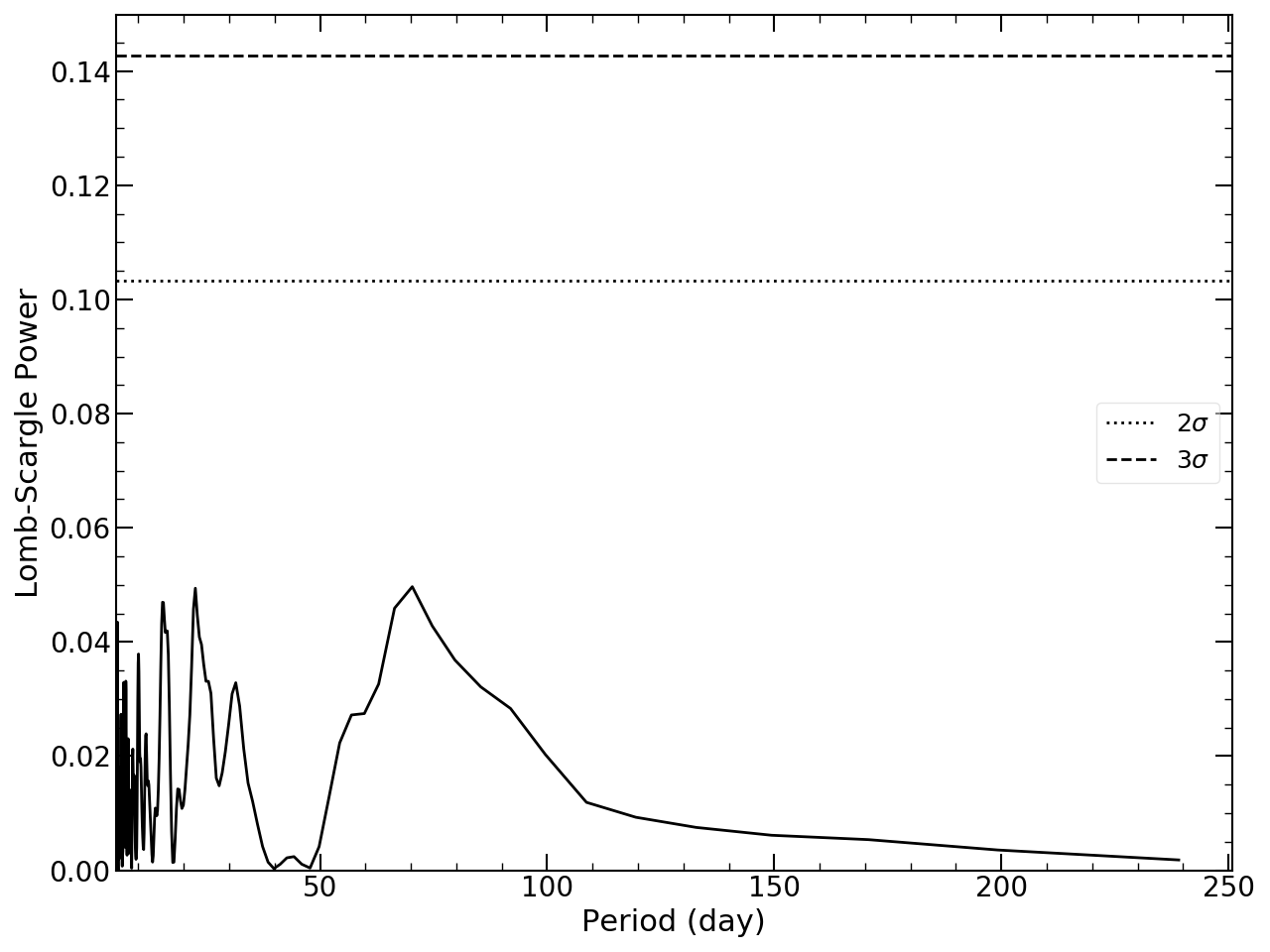}
     \caption{As per Figure \ref{fig:hoIIx1_periodogram_2009_2010}, but for the 2019 \swift-XRT dataset of \ho.}
     \label{fig:hoIIx1_periodogram_2019}
 \end{figure}

For \ho, in order to avoid aliasing effects caused by the large data gaps, we analysed the data separately in two segments of continuous monitoring. The first segment included the data from 2009 to 2010 and the second one comprised the 2019 data. The resulting periodograms are shown in Figures \ref{fig:hoIIx1_periodogram_2009_2010} and \ref{fig:hoIIx1_periodogram_2019}. In the inset of the Figures we also show the structure of the observing window, which is computed by performing the Lomb-Scargle periodogram on a light curve of constant flux and the same time coordinates as for the real data.
 The periodogram of the 2009--2010 segment reveals two strong peaks at around 21 and 28 days both with a significance above the 3$\sigma$ level. However, the observing window has a strong power at $\sim$ 25 days (Figure \ref{fig:hoIIx1_periodogram_2009_2010}), which may be introducing some aliasing. We also note that \cite{grise_x-ray_2010} analysed the same dataset and found no evidence of any periodicity. However, the discrepancy between these results might be simply due to the differences in the periodogram implementation used by the authors \citep[the prescription by][]{horne_prescription_1986}, the sampling grid, the way the false alarm probabilities were computed, and the fact that their implementation did not take into account the uncertainties in the measurements. 
 
The best solution given by the 28-day peak is shown in Figure \ref{fig:hoIIx1_periodogram_2009_2010} (see the light purple solid line in the top right panel). The variability is clearly more complex than described by the simple sinusoidal profile. It is also observed that the source variability changed after 2 February 2010 (see the blue dashed vertical line in the top right panel of Figure \ref{fig:hoIIx1_periodogram_2009_2010}). Since there is now compelling evidence of variable periodicities in ULXs \citep{kong_possible_2016, an_temporal_2016, brightman_60_2019,vasilopoulos_m51_2020} with an as yet unclear origin, we repeated our analysis by considering the data before and after 2 February 2010 separately. The results are presented in the top right and bottom panels of Figure \ref{fig:hoIIx1_periodogram_2009_2010}. The peak from the first segment is consistent with the one at 28 days previously found, albeit now is found slightly below the 3$\sigma$ level. For the second segment after the blue dashed line, we found no evidence of any periodicity above the 2$\sigma$ level. We still show the best solution from the highest peak found in the periodogram (yellow solid line in the top right panel of Figure \ref{fig:hoIIx1_periodogram_2009_2010}) to illustrate that if the oscillation is still present, its amplitude must have significantly decreased. Finally, the 2019 data showed no evidence of any periodicity, which may indicate that the oscillation has decreased significantly or disappeared entirely (see Figure \ref{fig:hoIIx1_periodogram_2019}). 
 
\begin{figure*}
    \centering
    \includegraphics[width=0.49\textwidth]{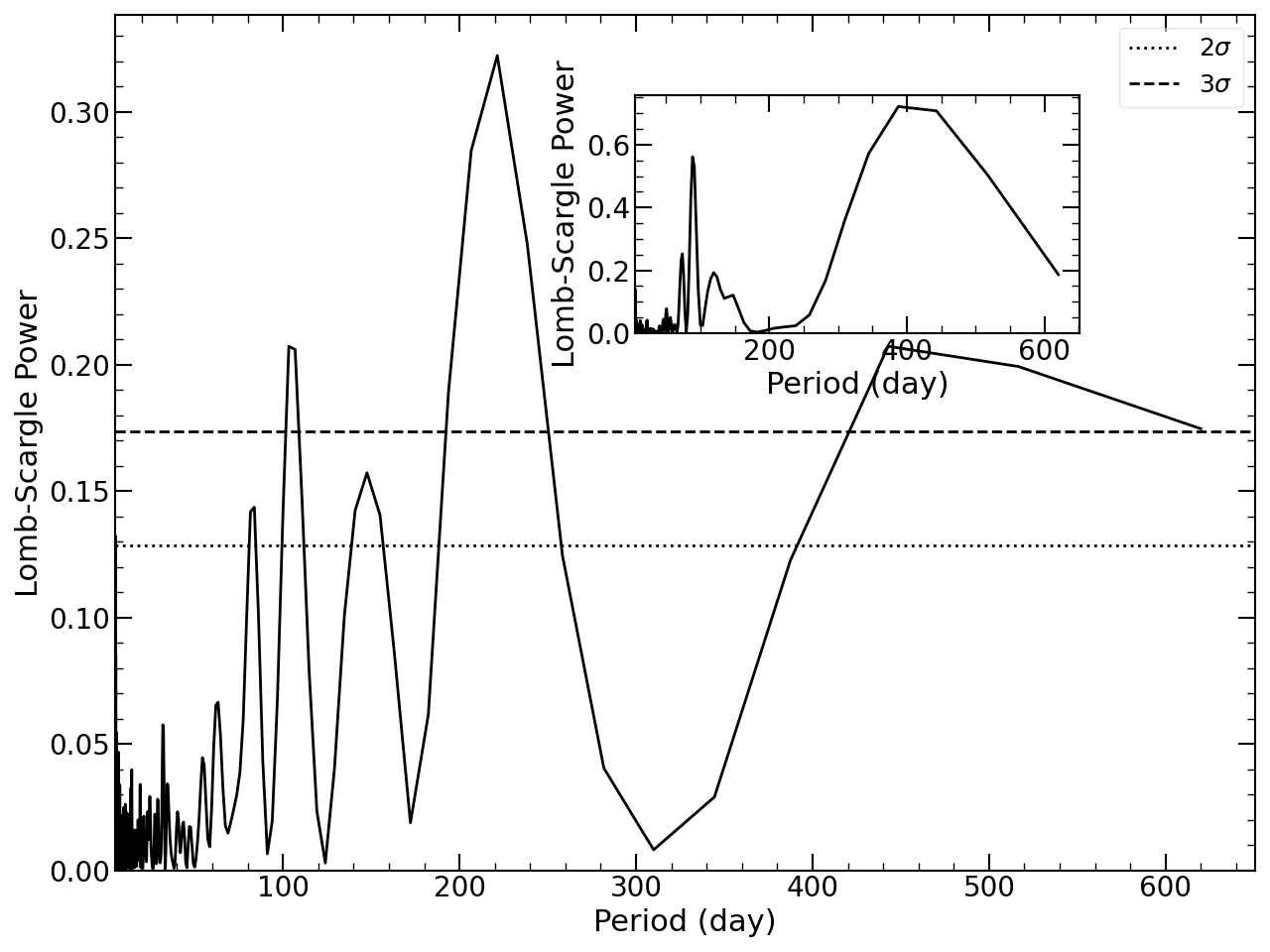}
    \includegraphics[width=0.49\textwidth, height=0.3\textheight]{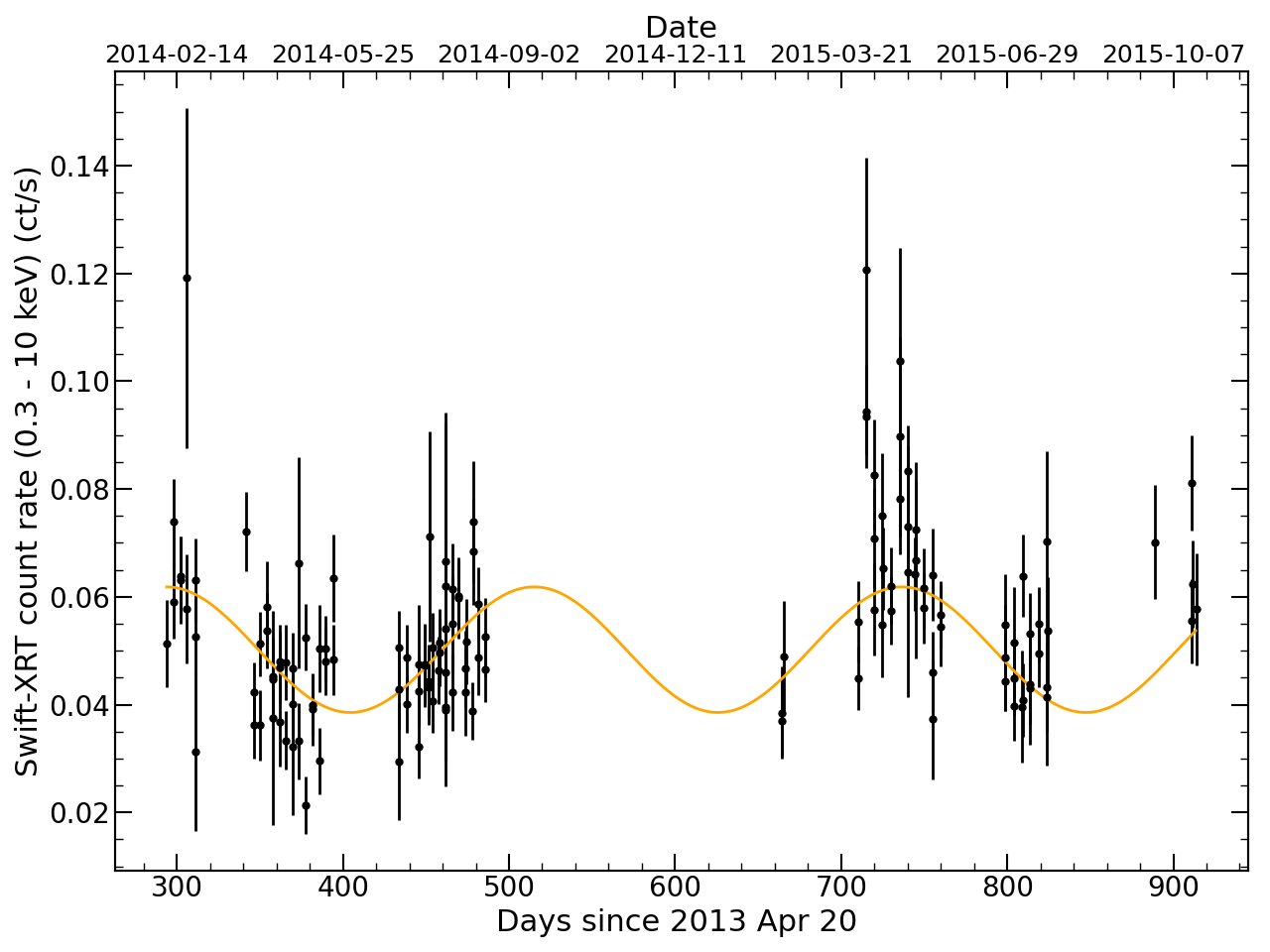}
    \caption{\textit{Left}: Lomb-Scargle periodogram of the 2014--2015 \swift-XRT light curve of \ngc\ in solid black. The 2$\sigma$ and 3$\sigma$ mark the false alarm probability levels computed from bootstrapped samples (see text for details). The inset plot shows the structure of the observing window. \textit{Right}: \swift-XRT light curve with the best-fit period from the periodogram (orange solid line).}
    \label{fig:periodicities_ngc5204x1}
\end{figure*}

For \ngc, we analysed only the continuous and densely sampled part of the light curve from $\sim$ day 300 until $\sim$ day 900 after the start of the monitoring (green and purple segments in Figure \ref{fig:swift_data}). The Lomb--Scargle periodogram reveals a strong peak at around 221 days (Figure \ref{fig:periodicities_ngc5204x1}), which is not coincident with any structure from the observing window. The peaks at 444 days and 103 days above the 3$\sigma$ are almost exactly or very close to 2 $\times$ 221 days and 221/2 days, respectively, suggesting that they could be harmonics of the period at 221 days. The peak at 444 days is also highly uncertain as it is only supported by 1.5 cycles, and stronger noise is expected towards longer periodicities given the sparsity of our data and the structure of the observing window. We also observe some variability on top of the sinusoidal variation at around 700--800 days after the beginning of the monitoring.

\subsubsection{False-alarm probability: REDFIT} \label{sec:redfit}
We note, however, that in the presence of aperiodic variability, the measurements could have some degree of correlation (i.e. in the presence of a flare) and by design, any correlation will be lost in the bootstrapped samples. In fact, red-noise, commonly observed in accreting sources, can often appear periodic \citep[e.g.][]{scargle_studies_1981, vaughan_false_2016}, and tests against the null-hypothesis of white noise often underestimate the false-alarm probability of the peaks in the periodogram \citep[e.g.][]{liu_no_2008}.

We therefore performed a second test to derive the significance of the peaks in the periodogram under the null-hypothesis of red noise. To do so, we based ourselves on the publicly available \texttt{R} code REDFIT developed by \cite{schulz_redfit_2002}\footnote{\url{https://rdrr.io/cran/dplR/src/R/redfit.R}}, in which red noise is modelled as a first-order autoregressive process (termed AR(1) for short), a model commonly used in astrophysics to model stochastic time series (see e.g. \citet{scargle_studies_1981} and also \citet{an_temporal_2016} for a similar approach). In AR(1) processes, the time series ($x_i$) at a given time ($t_i$) depend on the previous sample as $x_i$ = $\rho_ix_{i-1}$ + $\epsilon_i$, where $\rho$ is the autocorrelation coefficient and $\epsilon$ is the random component with zero mean and a given variance $\sigma^2$. The case in which $\rho$ = 1 simply describes a random walk. For unevenly sampled data, the AR(1) can be modelled with a variable autocorrelation coefficient that depends on the sampling time difference $\rho$ = $e^{-(t_i-t_{i-1})/\tau}$ \citep{robinson_estimation_1977}, where $\tau$ is the characteristic time of the AR(1) process, a measure of its memory and $\sigma^2$ is set equal to $1 - e^{-2(t_i-t_{i-1})/\tau}$ to ensure that the AR(1) process has a variance equal to unity and is stationary. $\tau$ can be calculated directly from the time series using the least-squares algorithm of \cite{mudelsee_tauest_2002}. From the AR(1) model, an ensemble of Monte Carlo simulations of time series with the same sampling as for the original light curve can be created to test whether the presence of a peak in the periodogram is consistent with an AR(1) process. However, the Lomb-Scargle implementation proposed by \cite{schulz_spectrum_1997} does not take into account the uncertainties in the data, and the mean of the measurements is not fitted as for the floating-mean periodogram. Additionally, REDFIT might underestimate the significance of the peaks in the periodogram as it computes the false alarm probabilities based on the entire distribution of the simulated periodograms rather than on the largest peak in each synthetic periodogram.

Because of the caveats outlined above, we replicated the same approach but instead simulated the ensemble of AR(1)-generated light curves by setting the variance of the noise $\sigma^2 = \sigma_d^2\,\left(1 - \mathrm{e}^{-2(t_i-t_{i-1})/\tau}\right)$, where $\sigma_d$ is the variance of the data, so that the simulated data has a variance equal to that of the real measurements. We then added an offset to the simulated data equal to the mean of the real measurements, so the simulated data also has approximately the same mean. We then computed uncertainties on $x_i$ assuming Poissonian statistics. In cases where $x_i$ happens to be negative, we simply set the uncertainty equal to that of the observed data at $i$. We note that there is no issue in introducing negative rates in the Lomb-Scargle calculation, although the simulations should ideally be flux limited to match the real measurements more closely.

\begin{figure*}
    \centering
    \includegraphics[width=0.48\textwidth]{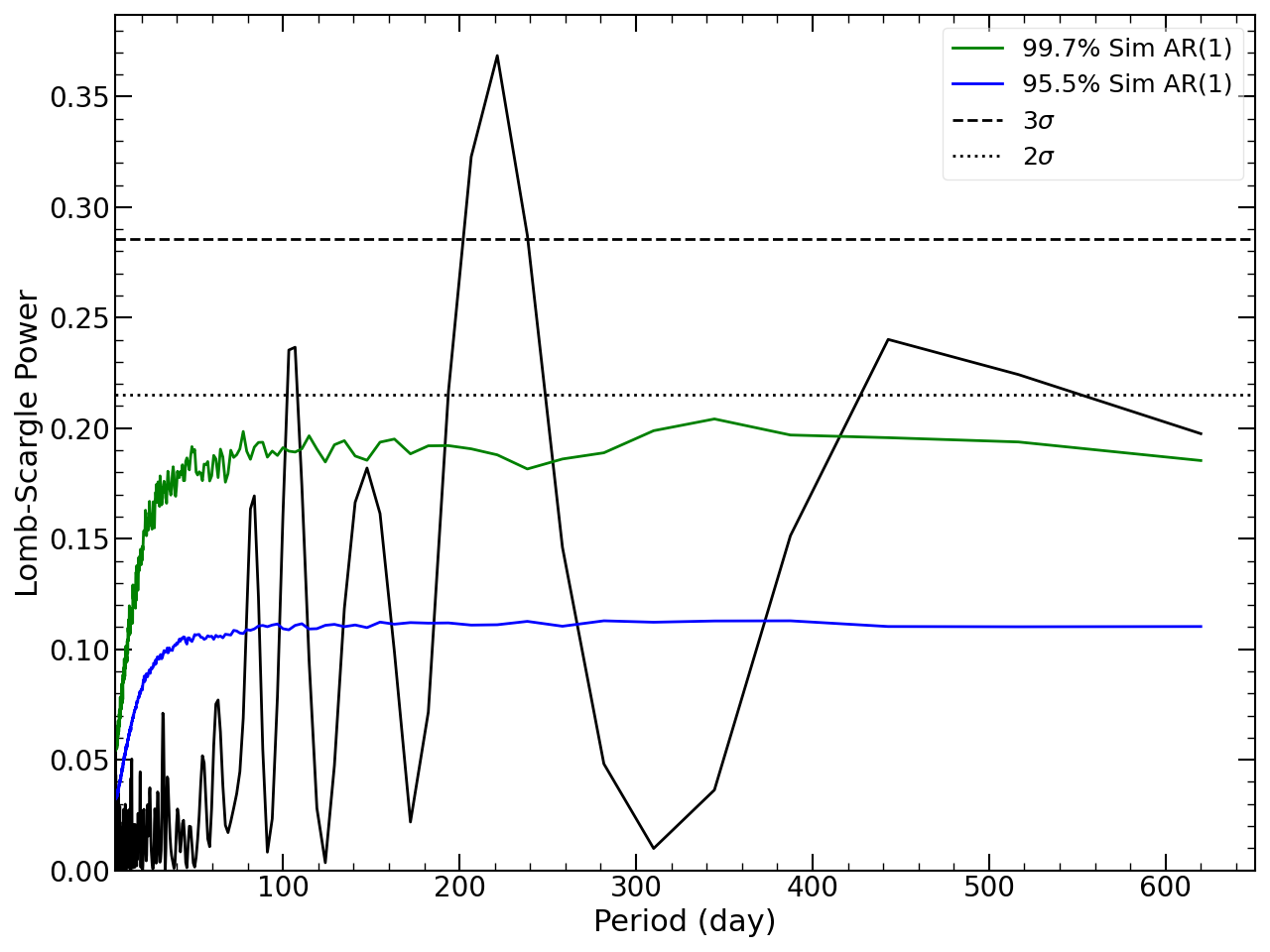}
    \includegraphics[width=0.48\textwidth]{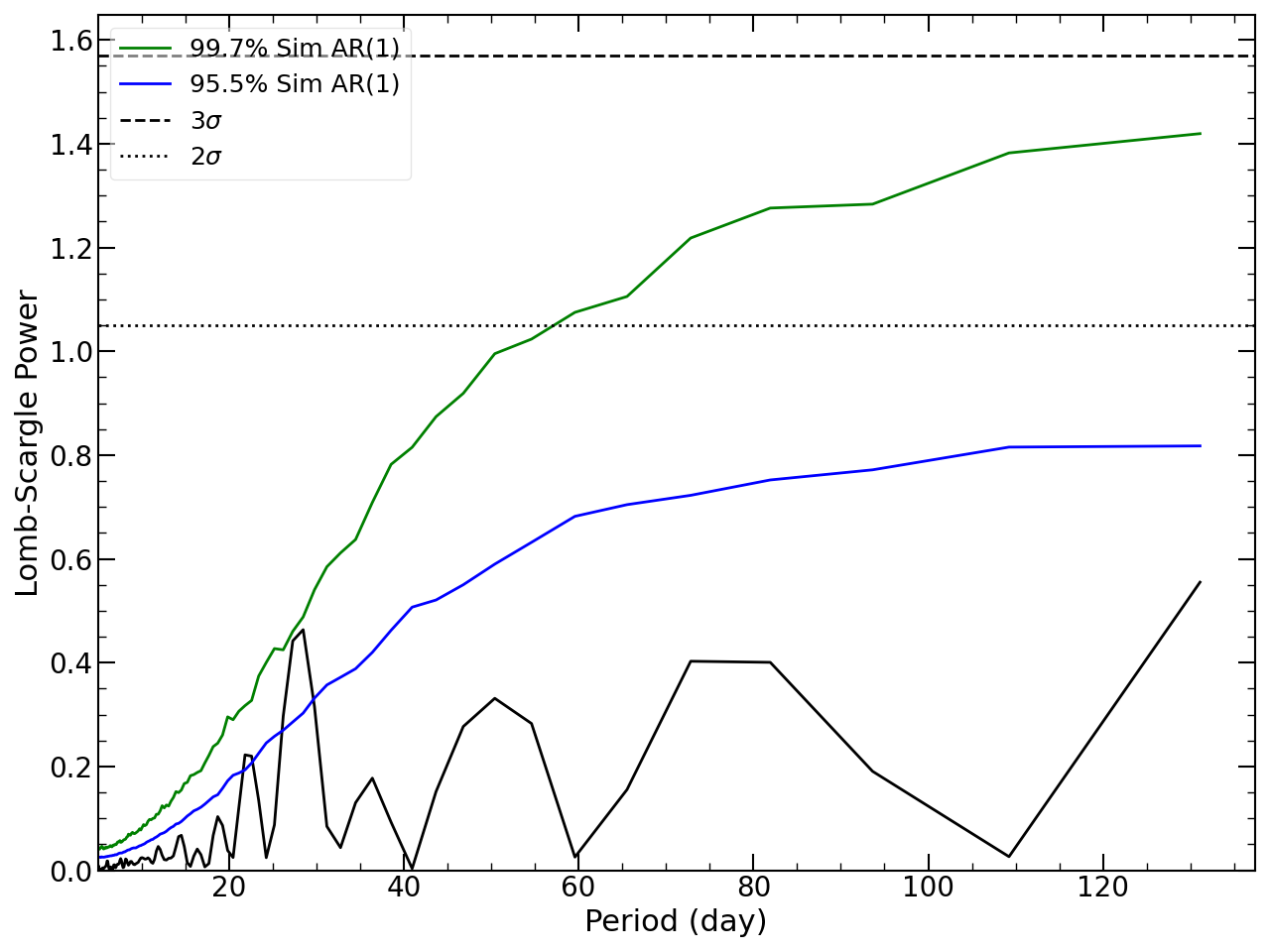}
    \caption{False-alarm probabilities for the Lomb-Scargle periodogram based on an adaptation of the REFIT code (see text for details) for \ngc\ (left) for the 2014--2015 monitoring and for \ho\ (right) for the 2009--2010 monitoring under the null-hypothesis of red noise. The coloured 95.5\% and 99.7\% mark the distribution of the 10000 synthetic periodograms, whereas the black dashed line marks 2$\sigma$ and 3$\sigma$ false-alarm probabilities, which have been determined from the maximum of the peaks of the synthetic periodograms.}
    \label{fig:custom_red_fit}
\end{figure*}

We then conducted the same Lomb-Scargle periodograms on these light curves as for the real data, taking into account the uncertainties. Following \cite{schulz_redfit_2002}, we scaled the synthetic periodograms to match the area under the power spectrum of the real data and computed a bias-corrected version of the periodograms. We finally derived the false-alarm probability levels from each of the largest peaks in the synthetic periodograms. Results are shown in Figure \ref{fig:custom_red_fit} for 10000 Monte Carlo simulations. 

As expected, the false alarm probabilities under the null-hypothesis of red noise are more stringent than those from the bootstrap method. For \ho,\ the peak is well below the 3$\sigma$ false-alarm probability. If we consider the less stringent false-alarm probability computed from the entire distribution of simulations (green line), the peak is only slightly below this limit. It is possible that the variability exhibited by the source (Figure \ref{fig:hoIIx1_periodogram_2009_2010}, right panel) is complicating the detection of the quasi-periodicity if the latter is not strictly coherent. Thus, more monitoring might be needed to clarify whether the variability is indeed quasi-periodic. Instead, for \ngc\ the red-noise AR(1) process cannot account for the variability observed, suggesting that the periodicity in the periodogram might be real.

\subsubsection{Uncertainties on the period}
In order to derive meaningful uncertainties on the periodicity of \ngc, we followed \citet{brightman_60_2019} and used a fitting procedure based on Gaussian process modelling proposed by \citet{foreman-mackey_fast_2017}. Gaussian processes offer several advantages over the classical Lomb--Scargle periodogram, namely the possibility to test more refined models than the implicitly assumed stable sinusoid in the Lomb--Scargle periodogram, and the possibility to retrieve meaningful errors on the best period estimate, which are not well defined in the case of the periodogram \citep{vanderplas_understanding_2018}. We used the python implementation \texttt{celerite} proposed by \citet{foreman-mackey_fast_2017} and considered three different models consisting of one or two components for our data.

The first model (model a) has a single term to capture the oscillatory behaviour by means of a damped simple harmonic oscillator (SHO), whose power spectral density is given by
\begin{equation} \label{eq:psd_sho}
    S(\omega) = \sqrt{\frac{2}{\pi}} \frac{S_0 \omega_0^4}{(\omega^2-\omega_0^2)^2 + \omega_0^2\omega^2/Q^2} 
,\end{equation}
where $S_0$ is the amplitude of the oscillation, $\omega_0$ is the frequency of the undamped oscillator, and $Q$ is the quality factor, which indicates how peaked the periodicity is in the PSD and gives a measure of the stability of the periodicity. For the second and third models, we added a noise component on top of the SHO to consider the variations not captured by the Lomb--Scargle solution (see e.g. Figure \ref{fig:periodicities_ngc5204x1} around 700--800 days after the beginning of the monitoring). In one of the models, this noise was modelled with a white noise or jitter term (model b) with amplitude $\sigma$. For the other model (model c), we considered a red-noise component that can be modelled in
\texttt{celerite} with the same kernel as for the damped harmonic oscillator, but fixing $Q_N$ = 1 / $\sqrt{2}$ \citep{foreman-mackey_fast_2017}, and leaving the other parameters of the oscillator ($S_N$, $\omega_N$) free to vary during fit. To summarise, we thus tested three different models: a) a single SHO term, b) the same SHO term but with a white noise term, and, finally, c) the SHO term with the red-noise component. 

In order to select the most suitable model for our datasets, we fitted each of the three models to the light curve using the L-BFGS-B minimisation method from \texttt{scipy} and retrieved the Bayesian information criterion (BIC) \citep{schwarz_estimating_1978} as given in Equation 54 of \citet{foreman-mackey_fast_2017}, meaning lower BIC values indicating preferred models. We set the following uniform priors on the parameters:10 $<$ $P$ $<$ 700 days, 0.1 $<$ $S_0$ $<$ 10000, and 20 $<$ $Q$ $<$ 2000. That is, all parameters are essentially unconstrained except for the lower bound of $Q$, which we limited, as low values of $Q$ lose the oscillatory behaviour that we want to describe \citep{foreman-mackey_fast_2017}. For the red-noise component, $S_N$ and $\omega_N$ are arbitrarily started at log $\omega_N$ = --0.1 and log $S_N$ = --5 with the following uniform priors (again essentially unconstrained): --10 $<$ log $\omega_N$ $<$ 15 and --10 $<$ log $S_0$ $<$ 15. The amplitude of the white noise term $\sigma$ is set initially to the mean value of the count-rate errors and is allowed to vary between --10 $<$ log $\sigma$ $<$ 15. We performed several runs varying the initial parameters to ensure that the minimisation routine finds the global minimum for each model. The best-fit parameters and the BIC calculations are presented in Table \ref{tab:bic}.

\begin{figure*}
    \centering
    \includegraphics[width=0.49\textwidth, height=0.35\textheight]{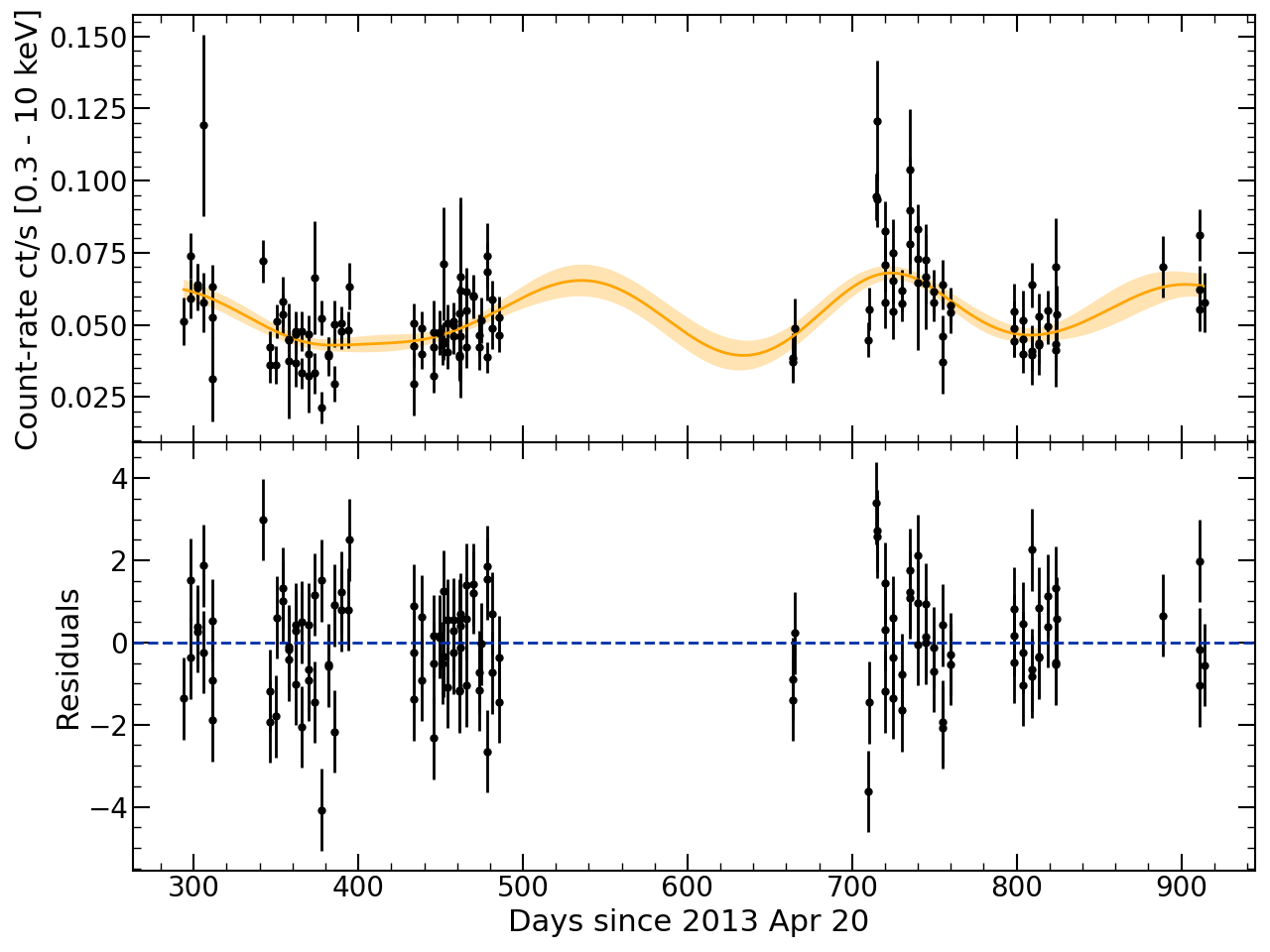}
    \includegraphics[width=0.48\textwidth, height=0.35\textheight]{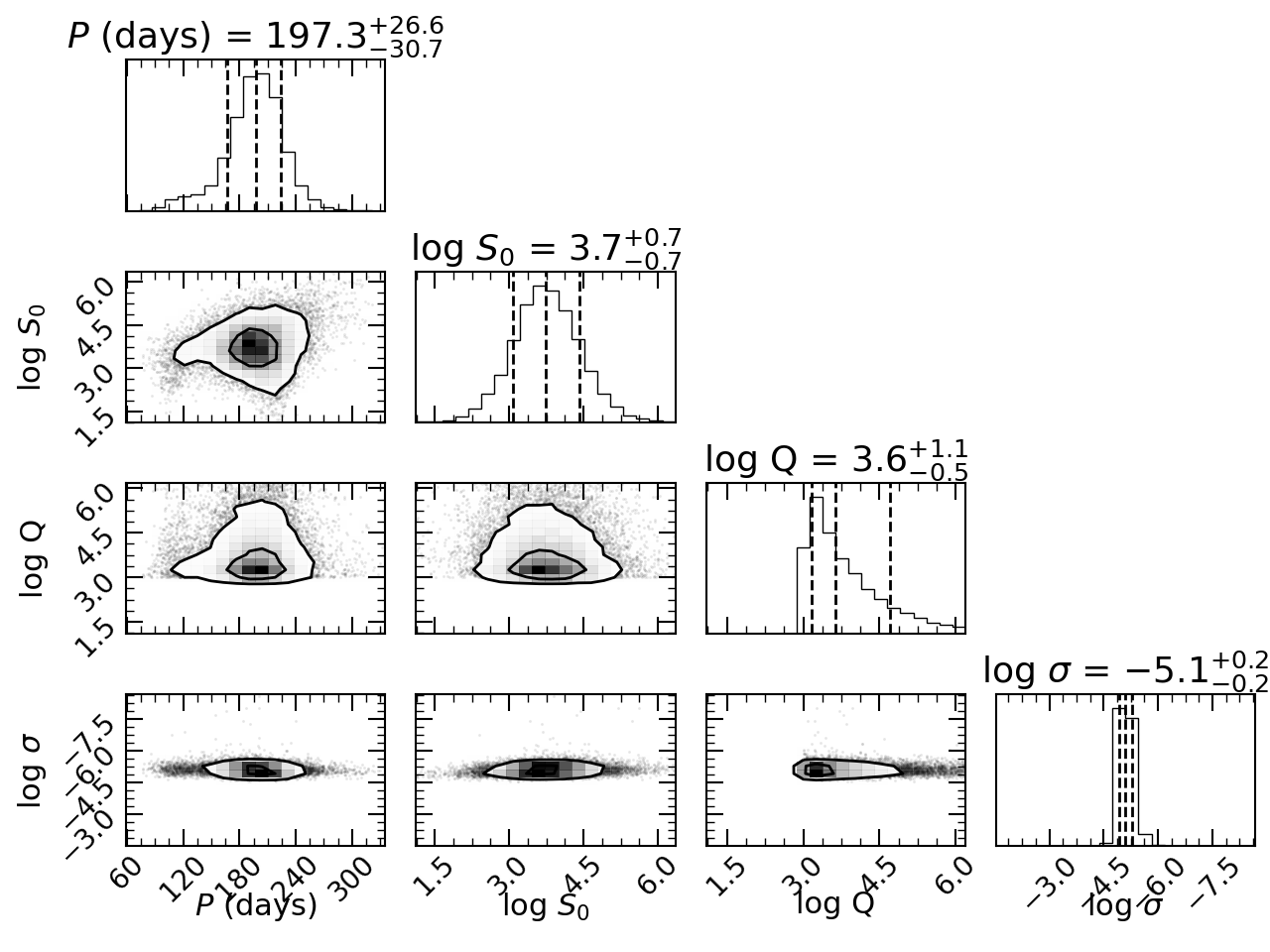}
    \caption{ \textit{Left:}  Best-fit solution of the \texttt{celerite} modelling (orange solid line) with the 1$\sigma$ shaded contours for the light curve of \ngc\. . The model shows the SHO with additional white noise (model b), for which we only show the oscillatory component. \textit{Right}: Posterior probability distributions of the parameters of the \texttt{celerite} model in the right panel. The histograms along the diagonal show the marginalised posterior distribution for each parameter with dashed lines indicating the median and the 1$\sigma$ errors. The two-dimensional histograms show the marginalised regions for each pair of parameters with the contours showing the 1 and 2$\sigma$ confidence intervals.}
    \label{fig:best_fit_ngc5204x1}
\end{figure*}

\begin{table*} 
 \centering 
 \caption{Best-fit parameters from the minimisation routine for the three different models tested and the corresponding Bayesian information criterion (BIC) for each of them to model the 2014--2015 light curve of \ngc.} 
 \label{tab:bic}
 \begin{tabular}{cccccccc} 
 \hline
 \noalign{\smallskip}
Model  & $P$ (days) & log $S_0$ & log $Q$ & log $\sigma$ &   log $S_\text{N}$ & log $\omega_\text{N}$ & BIC \\
\noalign{\smallskip}
\hline 
\hline
 \noalign{\smallskip}
\multicolumn{8}{c}{\ngc}\\
 \noalign{\smallskip}
SHO & 186.5 & 4.0 & 3.0 & - & - & -  &--772  \\
SHO + white noise & 196.8 & 3.7 & 3.0 & --5.1 &-&- &--788\\
SHO + red-noise & 196.8 & 3.7 & 3.0 & - & --6.9 &-2.9& --783 \\
\noalign{\smallskip}
\hline
 \end{tabular} 
 \end{table*}
Based on the BIC values, the model with the addition of white noise (model b) seems to be the preferred model. The addition of a more complex red-noise component (model c) seems not to be justified by the quality of data since a similar solution is found to that of model b (Table \ref{tab:bic}), with the addition of an extra parameter. Therefore, the simpler white noise component seems to be the preferred model, and we thus considered only this model for the subsequent analysis. The best-fit model is shown in Figure \ref{fig:best_fit_ngc5204x1}.

In order to estimate errors on the best fit parameters of the model with the added white noise, we initialised 32 walkers (or chains) to sample the posterior probability distribution by drawing 20000 MCMC samples using the \texttt{emcee} library in \texttt{python}. Each walker is initialised by randomly drawing a sample from the uniform priors defined above to ensure that parameter space is adequately sampled by the chains. In order to estimate an adequate burn-in period, we visually inspected the convergence of the chains and discarded the first 10000 samples of each of them. From the remaining samples, we selected one every 20 samples to derive the posterior probability density that is shown in Figure \ref{fig:best_fit_ngc5204x1}. The periodicity found above in the light curve of \ngc\ is of 197\errors{27}{31} days (1 $\sigma$ error), which is consistent with the 221-day peak found in the \ls\ periodogram (Section \ref{sec:redfit}) at the 1 $\sigma$ level.
       
\subsection{Spectral analysis}\label{sec:spectral_analysisi}
Here, we attempt to spectrally characterise each of the states found in Section \ref{sec:long_term_variability} (HUL, SUL, and SSUL regimes) with a physically motivated model. To do so, we made use of the \xmm, \chandra,\ and \nustar\ data used in \cite{gurpide_long-term_2021}, and the additional \chandra\ and \swift-XRT data of this work. We considered all the available datasets for each state, regardless of their date, and fitted them assuming the same model. This selection is driven by the results from Section \ref{sec:long_term_variability} and the analysis presented in \cite{gurpide_long-term_2021}. We also extracted \swift-XRT spectra for each source state by stacking individual observations using the online tools \citep{evans_methods_2009}. The count rates and hardness used to classify the observations can be found in Table \ref{tab:segregation}. For \ngc, we found the data quality of \swift-XRT spectra to be too low for a meaningful analysis. For \ho, we only found good agreement between the \swift-XRT data and the other mission datasets for the SUL regime, and thus this was the only \swift-XRT spectrum that we finally used. In the other cases, we found the \swift-XRT data to be significantly dimmer below 1 keV compared to the other datasets (residuals of $\gtrsim$ 3$\sigma$) and the floating cross-calibration constant is above 10\% with respect to the \xmm\ data, even when taking into uncertainties. This disagreement is consistent with the scatter seen in the \swift-XRT observations with respect to the \xmm\ observations of the SSUL and HUL regimes (Figure \ref{fig:swift_data}).
\begin{table} 
 \centering 
 \caption{\swift-XRT count rates and hardness ratios used to extract stacked spectra.}\label{tab:segregation} 
 \begin{tabular}{lcccc} \noalign{\smallskip}
 \hline \hline
 \noalign{\smallskip}
& \multicolumn{2}{c}{\ngc} & \multicolumn{2}{c}{Holmberg II X-1} \\ 
\cline{2-3}\cline{4-5}
  \noalign{\smallskip}
State & Count rate & HR & Count rate & HR \\
 & cts/s &  & cts/s &  \\
\noalign{\smallskip}
\hline
\noalign{\smallskip}
SUL & 0.06--0.11 & 0.4--0.8&0.225--0.35& 0.3--0.7\\
HUL & 0.05--0.06 & 0.45--1.1&0.105--0.205& 0.44--1.0\\
SSUL  & 0.4--1.1 & 0.035--0.05&0--0.105& 0--0.8\\
  \noalign{\smallskip}
  \hline
 \end{tabular}
  \end{table}
  
We also inspected the \xmm\ optical monitor and the \swift-UVOT images looking for observations that could help to constrain the broadband emission. However, given the limited spatial resolution of the instruments (PSF $\sim$ 1 arcsec), both ULXs are completely blended with nearby sources in the field of view preventing a clean determination of the direct contribution from the ULX. We therefore searched for \hubble\ optical/UV data using the MAST portal\footnote{\url{https://mast.stsci.edu/portal/Mashup/Clients/Mast/Portal.html}} and selected observations when the source state was known. We restricted the search to the wide or medium filters as we were interested in the continuum emission. Unfortunately, we only found quasi-simultaneous \hst\ data for the hard-intermediate state of Holmberg II X-1 in the F275W, F336W, F438W, and F550M filters. We used a 0.255" radius circular aperture for the source and an annulus of 0.37" and 0.58" inner and outer radius for the background, to consider some of the nebular emission in the background subtraction. We applied the appropriate aperture correction for each filter \footnote{\url{https://www.stsci.edu/hst/instrumentation/acs/data-analysis/aperture-corrections} and \url{https://www.stsci.edu/hst/instrumentation/wfc3/data-analysis/photometric-calibration/uvis-encircled-energy}} and converted the telescope counts to fluxes using the calibration \textit{PHOTFLAM} keywords found in the file headers\footnote{\url{https://www.stsci.edu/hst/wfpc2/Wfpc2_dhb/wfpc2_ch52.html}}. We obtained background-corrected fluxes in units of erg/cm$^{-2}$s$^{-1}$\AA$^{-1}$ of (6 $\pm$ 2) $\times$ 10$^{-17}$ ,  (3.6 $\pm$ 0.8) $\times$ 10$^{-17}$, (1.6 $\pm$ 0.4) $\times$ 10$^{-17}$ , and (7$\pm$ 2) $\times$ 10$^{-18}$ for the F275W, F336W, F438W, and F550M filters, respectively. The different datasets considered together to characterise each state can be found in Table \ref{tab:states}.

Previous broadband spectral analysis by \cite{tao_nature_2012} of \ngc\ indicated that the emission could be consistent with an irradiated standard accretion disc and that a stellar origin was unlikely. Here, we chose to model our data with the physically motivated model \texttt{sirf} \citep[][available in \xspec]{abolmasov_optically_2009}, which provides a self-consistent analytical solution to the super-critical funnel geometry based on the theoretical works of \cite{shakura_black_1973} and \cite{ poutanen_supercritically_2007} to see if the broadband emission could instead be explained as arising from a self-irradiated super-critical disc. 

This model considers black-body emission from the outer wind photosphere, the walls of the super-critical funnel, and the photosphere at the bottom of the funnel, taking into account self-irradiation effects. The model has nine variable parameters: namely, the temperature and the distance from the source of the funnel bottom photosphere ($T_\text{in}$ and $R_\text{in}$), the outer wind photospheric radius ($R_\text{out}$), the half-opening angle of the funnel (\afunnel), the velocity-law exponent for the wind ($\alpha$) that is, $v_\text{wind}$ $\propto$ $r^\alpha$, the adiabatic index ($\gamma$), the mass-ejection rate (\mout), and the normalisation of the model. As a caveat, we note that the model does not take into account Comptonisation, which is likely to be an important source of opacity in the super-Eddington regime \citep[e.g.][]{kawashima_comptonized_2012}.
 
We modelled interstellar X-ray absorption with two \texttt{\tbabs} components, one frozen at the Galactic value along the line of sight \citep[2.72 $\times$ 10$^{20}$ cm$^{-2}$ and 5.7 $\times $10$^{20}$ cm$^{-2}$ for \ngc\ and Holmberg II X-1, respectively;][]{hi4pi_collaboration_hi4pi_2016} and another one free to vary. For the dataset containing \hst\ data, we considered interstellar extinction by adding the \texttt{redden} model in \texttt{XSPEC} with E(V-B) fixed to 0.074, based on the measured $H_\alpha$/$H_\beta$ ratio from the nebular emission of \cite{vinokurov_ultra-luminous_2013} and using the extinction curve from \cite{cardelli_relationship_1989} of $R_V$ = 3.1.

The nine parameters of the \texttt{sirf} model could not all be constrained if left free to vary, even for the hard-intermediate state of \ho, where we had the best data quality. Therefore, following \cite{abolmasov_optically_2009}, in all fits we fixed the velocity law exponent of the wind ($\alpha$) to --0.5, the adiabatic index ($\gamma$) to 4/3, and R$_\text{out}$ to 100 $R_\text{sph}$, as the latter parameter is insensitive to the spectra unless its value is below 1, according to the authors. However, we found that for the soft and super-soft states, due to the lack of optical coverage and the poor data quality, we could not discriminate between different solutions, with the fit often being insensitive to several parameters and resulting in large parameter uncertainties. Another complication arose due to the fact that the model favours solutions in which $i$ $\sim$ \afunnel\ when \afunnel\ was not well constrained. This occurs due to the fact that a sharp transition is created in the $\chi^2$ space when the model changes from a solution where our line of sight is within the the wind cone to a solution with the wind cone out of the line of sight, creating an artificial minimum in the $\chi^2$ space at $i$ $\sim$ \afunnel. We therefore decided to restrict the spectral analysis to the hard-intermediate states, which are the only states for which we have good quality broadband data. Obtaining \hst\ observations of both ULX in the different spectral states should help to allow us to constrain the broadband emission.
 
 \begin{table*} 
 \centering 
 \caption{Best-fit results for the HUL states seen in \ngc\ and \ho\ using the \texttt{sirf} model. All uncertainties are given at the 90\% confidence level.} \label{tab:fit_states}
 \begin{tabular}{lcccccccc} 
 \hline\hline
  \noalign{\smallskip} 
 State & \nh\ & $T_\text{in}$ & $R_\text{in}$ & \afunnel & $i$ & \mout\ & norm  & \chisqr\ / dof\\ 
  & 10$^{20}$ cm$^{-2}$ & keV  & 10$^{-3}$ $R_\text{sph}$ & $^\circ$ & $^\circ$  &  &  & \\
     \noalign{\smallskip} 
 \hline \hline 
 \noalign{\smallskip}
 \multicolumn{9}{c}{\ngc} \\ 
 \noalign{\smallskip} 
HUL & 7$\pm$1 & 1.45\errors{0.12}{0.09}& 0.8\errors{0.3}{0.2}& 39.4$\pm$0.4& $<$ 12.2 & 29\errors{10}{4}& 4.9\errors{4.3}{2.2}& 1.12 (1136) \\
   \noalign{\smallskip} 
 \hline 
  \noalign{\smallskip} 
\multicolumn{9}{c}{\ho} \\ 
 \noalign{\smallskip}
HUL & 11.4$\pm$0.7 & 1.48\errors{0.05}{0.04}& 0.15\errors{0.04}{0.01}& 41.0$\pm$0.4& $<$ 9.1 & 169$\pm$25& 544.2\errors{183.1}{157.6}& 0.99 (960) \\ 
  \noalign{\smallskip}
   \hline 
 \hline 
 \noalign{\smallskip}
  \end{tabular} 
  \tablefoot{ All uncertainties are given at the 90\% confidence level. We set a lower limit on $i$ of 0.5$^\circ$.}
\end{table*}
 
\begin{figure*}
     \centering
     \includegraphics{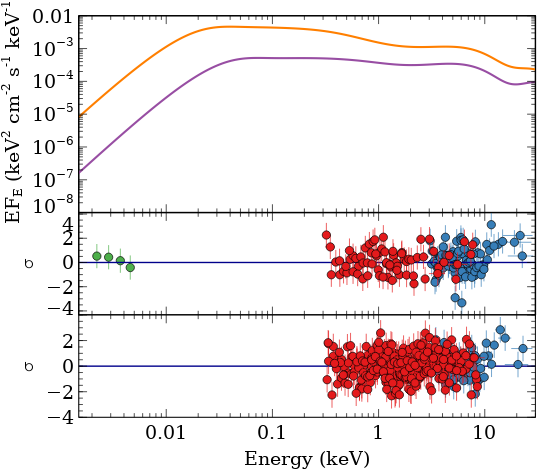}
     \caption{(Top) Broadband best-fit (solid lines) to the HUL states of Holmberg II X-1 (orange) and \ngc\ (purple line) with the self-irradiated super-critical funnel model (\texttt{sirf} in \xspec) for $i$ $\sim$ 10 degrees (see Table \ref{tab:fit_states}). The models in the upper panel have been corrected for X-ray and optical absorption. The lower panels show the best-fit residuals for \ho\ (middle panel) and \ngc\ (bottom panel) without any correction for X-ray absorption. Green, red, and blue circles show the \hst, \xmm\ EPIC-pn, and FPMA \nustar\ data, respectively (see Table \ref{tab:fit_states})}
     \label{fig:broadband_fit}
 \end{figure*}

 The \texttt{sirf} model offered a good fit (\chisqr = 1.00 for 960 degrees of freedom) to the broadband emission (see Figure \ref{fig:broadband_fit} and Table \ref{tab:fit_states}) of \ho\ in the HUL state. Thus, while emission from the donor star or an irradiated disc could be consistent with the UV/optical emission \citep[e.g.][]{tao_nature_2012}, self-irradiation by a super-critical disc seems another plausible scenario based on the broadband emission. We found an upper limit on $i$ of 9.1$^\circ$. This is slightly below the lower limit of 10$^\circ$ suggested by \cite{cseh_unveiling_2014}, assuming that the radio jet detection is not strongly Doppler boosted, although we note that if the source precesses, the viewing angle could be variable. Therefore, a viewing angle looking down the funnel (\afunnel $\sim$ 40 degrees) seems to be supported. For \ngc, we similarly obtained an upper limit on $i$ of $\sim$ 12$^\circ$. Interestingly, the set of parameters obtained are very similar to those obtained for \ho, except for a factor of $\sim$ 7 lower \mout\ and a factor of $\sim$ 5 higher $R_\text{in}$ (see Table \ref{tab:fit_states}). Therefore, the spectral differences between \ngc\ and \ho\ could be accounted for by a lower \mout\ and a thicker bottom of the funnel in the case of \ngc\ with an unobscured view of the bottom of the funnel in both cases. 
\section{Discussion} \label{sec:discussion}
The multi-instrument data analysed in this work has allowed us to firmly establish, for the first time, the presence of a recurrent evolutionary cycle in ULXs. The similarities that exist in the temporal (Figures \ref{fig:swift_data} and \ref{fig:hld_ngc5204x1}) and spectral \citep[Figure \ref{fig:broadband_fit} and Table \ref{tab:fit_states}; see also Figure 6 in ][]{gurpide_long-term_2021} properties of \ho\ and \ngc\ allow us to describe the evolution of the cycle as follows: starting from the HUL state, the sources transit to the SUL state (Figure \ref{fig:hld_ngc5204x1}). There, the sources will transit back and forth from the SUL regime to the SSUL regime (Figures \ref{fig:swift_data} and \ref{fig:hld_ngc5204x1}), and finally the sources will return to the HUL state from the SUL state (Figure \ref{fig:hld_ngc5204x1}). These findings offer a unique opportunity to constrain the timescale and nature of the transitions frequently observed in these sources \citep[e.g.][]{roberts_chandra_2006, grise_x-ray_2010} and could potentially be extended to similar transitions seen in other ULXs as well \citep[e.g.][]{middleton_diagnosing_2015, gurpide_long-term_2021, mondal_spectral_2021}. 

Spectral transitions in ULXs have been frequently discussed invoking geometrical effects induced by the wind/funnel structure \citep[e.g.][]{sutton_ultraluminous_2013, middleton_spectral-timing_2015, pinto_xmm-newton_2021} expected to form as the mass-transfer rate approaches the Eddington limit and radiation pressure starts to drive a conical outflow \citep{shakura_black_1973, poutanen_supercritically_2007}. Building a coherent picture of the long-term spectral evolution of these two ULXs was attempted in \cite{gurpide_long-term_2021}, where it was suggested that the transition from the hard to the soft ULX regime was due to an increase of the mass-transfer rate and corresponding decrease of the opening angle of the funnel, based on the HUL-SUL softening from a rather sparse monitoring provided by \xmm\ and \chandra. The authors also suggested that a line of sight grazing the wind walls in the SUL state could explain the rapid transitions observed to the SSUL state as we switch between peering down the optically thin funnel and observing the wind photosphere as the source precessed \citep[e.g.][]{abolmasov_optically_2009}. We note that a viewing angle grazing the wind walls is also supported by the lack of short-term variability in the SUL state of \ho\ and \ngc\ \citep[e.g.][]{heil_systematic_2009, sutton_ultraluminous_2013}, which is at odds with the high short-term variability seen in softer of the soft ULXs \citep[e.g.][]{middleton_challenging_2011} argued to be viewed through the wind. However, the lack of periodic variability (Figure \ref{fig:custom_red_fit}) in the SUL--SSUL transitions found here suggests that precession of the accretion flow might not be responsible for the spectral changes. Instead, a scenario in which the short-term variability may be ascribed to the presence of wind clumps crossing our line of sight \citep{takeuchi_clumpy_2013, middleton_spectral-timing_2015} as a result of the narrower funnel is supported by the aperiodic transitions. 

Similar transitions from SUL to SSUL are seen in the ULS NGC 247 ULX--1 \citep{urquhart_optically_2016, feng_nature_2016} and NGC 55 ULX \citep{pinto_ultraluminous_2017}, which are also observed during the bright states \citep{pinto_xmm-newton_2021}, suggesting a tight link between standard ULXs and ULSs. However, as opposed to NGC 247 ULX-1, \ho\ and \ngc\ are the first sources observed to switch through all these three canonical ULX states, possibly due to the lower viewing angle compared to ULSs. Moreover, dips frequently seen in ULSs \citep[e.g.][]{stobbart_xmmnewton_2006,urquhart_optically_2016, alston_quasi-periodic_2021} are not observed in the light curves of \ho\ and \ngc, supporting this interpretation. It is thus likely that the dipping activity in ULSs is associated with direct obscuration of the inflated accretion disc as a result of the increased mass-accretion rate \citep{guo_thick-disc_2019}. Hence, the increase in mass-accretion rate confers \ho\ and \ngc\ with a somewhat harder ULS aspect, because the SSUL aspect of these standard ULXs is not due to a high inclination angle, but rather due to the extreme narrowing of the funnel. This might imply that ULSs are characterised by both a high inclination angle and a high mass-accretion rate. 

\begin{figure*}
    \centering
    \includegraphics[width=0.6\textwidth]{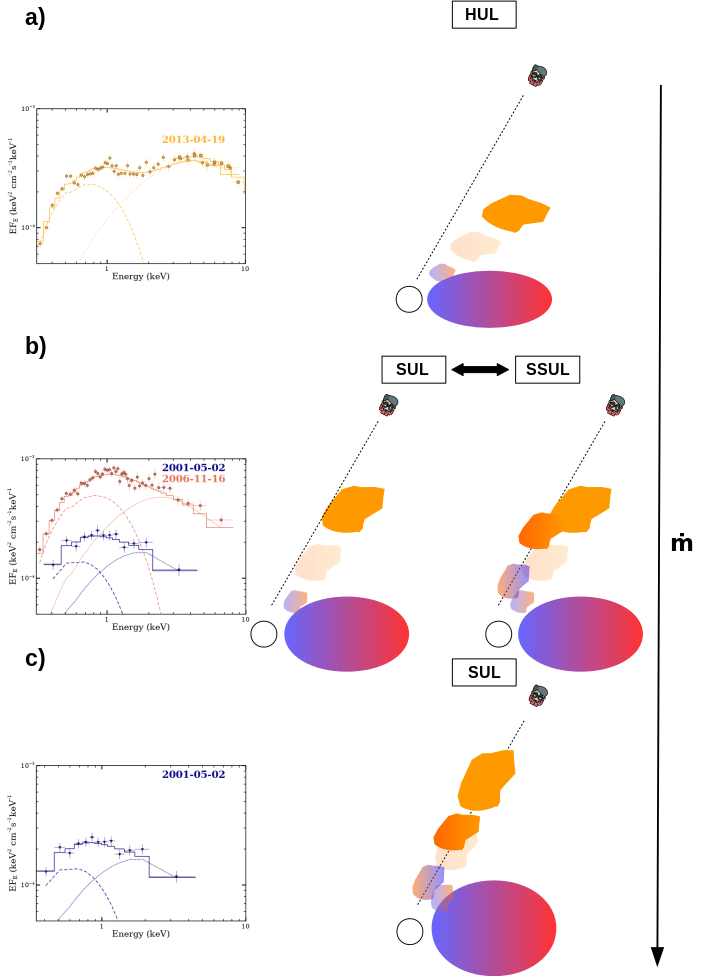}
    \caption{Cartoon illustrating the different spectral states observed in \ho\ and \ngc\ as a function of the mass-transfer rate and wind clumps crossing the line of sight. The compact object (NS or a BH) is represented by a white sphere. From top to bottom, the mass transfer increases as indicated by the black arrow on the right. In panel a), the funnel is wide open due to the low accretion rate, and the observer sees the inner regions of the accretion flow and little variability is observed (HUL regime; 2019 \swift\ monitoring of \ho), because the wind is unlikely to enter the line of sight. The left graph shows the corresponding spectral state \citep[adapted from][]{gurpide_long-term_2021}. In panel b), the mass-transfer rate increases and narrows the opening angle of the funnel (SUL  regime; SSUL regime), so wind clumps might now start to enter the line of sight and produce the spectral changes from soft to the SSUL regime observed (right and left panels and as seen in the 2009 \swift\ monitoring of \ho). Finally, in panel c), the mass-transfer rate increases further and the wind cone is now in the line of sight. The observer mostly sees the wind and the colder outer parts of the inflated disc, so the effect of absorption by wind clumps is dampened (end of the 2009--2010 \swift\ monitoring of \ho).}
    \label{fig:cartoon}
\end{figure*}

The amplitude of the jittering decays towards the end of the 2009--2010 monitoring of \ho\ (Figure \ref{fig:hoIIx1_periodogram_2009_2010}). This might imply that the mass-transfer rate has increased and narrowed the funnel further; so, towards the end of the 2009--2010 monitoring our line of sight no longer sees the inner regions of the accretion flow and instead mostly sees the wind photosphere, reducing the observed variability. Such dimming and reduced variability caused by an increase of the mass-accretion rate is in good agreement with the predictions made by \cite{middleton_spectral-timing_2015} for sources at moderate inclinations. This is illustrated in Figure \ref{fig:cartoon}, panels b and c.

The spectral analysis might support a low viewing angle for \ho\ and \ngc,\ at least for the HUL state (Section \ref{sec:spectral_analysisi}), and a moderate opening angle of the funnel (\afunnel\ $\sim$ 40$^\circ$). The lack of clear jittering behaviour in the 2019 data (see Figure \ref{fig:swift_data}) may be explained by the larger opening angle of the funnel as a result of the lower mass-transfer rate, implying that the wind is unlikely to enter the line of sight. This is consistent with the reduced short-term variability seen in hard ULXs \citep[e.g.][]{middleton_spectral-timing_2015, sutton_ultraluminous_2013}, frequently argued to be viewed face-on, and is consistent with the suggestion that the increase of mass-transfer rate and narrowing of the opening angle of the funnel causes the HUL--SUL spectral change (Figure \ref{fig:hld_ngc5204x1}). This interpretation naturally accounts for the lack of direct transitions from the SSUL regime to the HUL regime (Figures \ref{fig:swift_data} and \ref{fig:hld_ngc5204x1}). A cartoon illustrating the source variability and its relation to the different spectral states is depicted in Figure \ref{fig:cartoon}. 

The $\sim$ 200-day periodicity found in the light curve of \ngc\ is possibly associated with the timescale of the whole cycle (Figure \ref{fig:best_fit_ngc5204x1}); albeit, further investigation is needed. While the nature of this periodicity remains unclear, should our interpretation of the evolutionary cycle be correct, we would expect some mechanism of periodic mass-transfer increase to be responsible for it and cause the source to switch from the HUL regime to the SUL state. This could be the case in an eccentric orbit in which the mass-transfer rate is expected to be maximal as the companion passes through the periapsis. This scenario would likely require a super-giant star, as for large eccentricities $e$ $>$ 0.5 we find that the star would need to have a radius in excess of $\sim$ 200 $R_\odot$ to fill its Roche-Lobe overflow for compact objects of masses as low as 1.4 \msun. This large radius is at odds with the suggested radius of 30 $R_\odot$ for the companion based on the UV spectrum of the counterpart \citep{liu_optical_2004}. Additionally, orbital solutions in ULXs (only available thanks to a timing solution in ULXs showing pulsations) require modest eccentric orbits \citep[$e$ $\lesssim$ 0.1][]{bachetti_ultraluminous_2014, furst_tale_2018, rodriguez-castillo_discovery_2020} and generally shorter orbital periods ($P$ $\lesssim$ 60 days) are found, although the orbital period of NGC 300 ULX1 is expected to be of $\sim$ 1 year based on the red super-giant companion \citep{heida_discovery_2019}. 

Another mechanism that could induce periodic increases in the mass-transfer rate is the Kozai mechanism \citep{kozai_secular_1962}, in which the presence of a third body causes modulation of the eccentricity of the inner binary. The changes in eccentricity will affect the distance between the inner L1 Lagrangian point and the centre of mass of the companion, therefore modulating the mass-transfer rate. This mechanism has been proposed to explain the super-orbital X-ray modulation of $\sim$ 176 days associated with mass-transfer rate variations in the triple system 4U1820-303 (see \citet{zdziarski_superorbital_2007} and references therein). It has also been suggested that this mechanism could explain the $\sim$ 60-day quasi-periodicity responsible for the strong bi-modal flux modulation of the PULX M82 X--2 \citep{brightman_60_2019, bachetti_all_2020}, which has been argued to be due to the NS switching back and forth from the accretor phase to the propeller regime \citep{tsygankov_propeller_2016}. 

It is also worth noting that a linear relationship has been found between super-orbital and orbital periods in Be/X-ray binaries, SGXBs, and ULXs \citep[see][and references therein]{corbet_superorbital_2013, townsend_orbital_2020}. Assuming the 200-day periodicity is super-orbital in nature, we would expect an orbital period in \ngc\ of the order of $\sim$ 10 days. This is in agreement with the estimate made by \cite{liu_optical_2004}, based on the UV spectrum of the counterpart and on the assumption that the companion fills its Roche Lobe. Indeed, the relationship between the orbital and super-obital periods would imply that \ngc\ is a system with a super-giant filling its Roche Lobe \citep{townsend_orbital_2020}. We note that the linear relationship between the period and super-orbital period in super-giant Roche-Lobe filling systems is yet unclear, although the presence of a third body could account for the super-orbital periodicity \citep{corbet_superorbital_2013}, as stated also above.

Finally, we note that further monitoring of the soft spectral transitions would be of interest to robustly determine whether the spectral changes are associated with a quasi-periodicity. It has recently been proposed that a possible mechanism to account for the observed superorbital periods in ULXs could be the Lense-Thirring effect \citep{middleton_lense-thirring_2018}. If so, precession of the accretion flow induced by the Lense-Thirring effect may explain the spectral changes from SUL to SSUL as the wind cone comes into and out of the line of sight throughout the precession cycle, strongly modulating the flux \citep{abolmasov_optically_2009, dauser_modelling_2017}. The decay of the amplitude of the putative modulation towards the end of 2009--2010 could be similarly accounted for by an increase in the mass-accretion rate associated with a reduction of the opening angle of the funnel, meaning that throughout the whole cycle we only see the wind. Conversely, the lack of modulation during the HUL state of 2019 (Figure \ref{fig:hoIIx1_periodogram_2019}) could be due to the opposite effect, that is, a larger opening angle of the funnel resulting from a decrease of the
mass-transfer rate. In this case, our line of sight observes the optically thin funnel \citep[e.g.][]{narayan_spectra_2017} throughout the whole precessing cycle, again effectively diminishing the amplitude of the oscillation \citep{abolmasov_optically_2009}.

\section{Conclusions} \label{sec:conclusions}
We show the striking similarities of \ho\ and \ngc\ in their long-term evolution as they transit through the canonical hard, soft, and super-soft ULX spectral regimes. This is the first time that such recurrent evolutionary cycle is observed in ULXs and that all three states are observed in a single ULX, strengthening the hypothesis that ULSs are standard ULXs viewed at high inclinations. We argue that the sources transit from hard to soft ULX as the mass-transfer rate increases and the funnel closes. There, the source will exhibit jittering, as wind clumps start to cross the line of sight due to the now narrower funnel, explaining the spectral transitions from soft to super-soft ULX. As the mass-accretion rate increases further and the funnel closes, our line of sight mostly sees the wind photosphere, with the subsequent reduction of short-term variability. The absence of SSUL--HUL state transitions are naturally explained by this scenario. 

This interpretation is supported by the \swift-XRT light curve of \ho, which shows a lack of strong variability in the hard ULX state and a strong rapid variability responsible for the spectral changes between soft and super-soft states. The light curve of \ngc\ shows instead a periodicity on a long timescale of $\sim$ 200 days, which is likely associated with the duration of the full hard---soft--super-soft cycle; albeit, further monitoring is needed to robustly confirm this association. We argue that the transitions from hard to soft ULX are likely associated with some periodic increase of the mass-transfer rate, yet the nature of this modulation remains unclear. Similar periodicity may be present in the long-term evolution of \ho\ based on its similarities with \ngc. We therefore encourage \swift-XRT monitoring of \ho\ over longer timescales to detect it.

\begin{acknowledgements}
We would like to thank the anonymous reviewer for comments and
suggestions that improved this manuscript. The authors are grateful to H. Carfantan for his help with the timing analysis. GV acknowledges support by NASA Grant numbers 80NSSC20K1107, 80NSSC20K0803 and 80NSSC21K0213. NAW acknowledges support by the CNES.\\
Software used: HEASoft v6.26.1,  Python v3.8, Veusz 3.3.1 (J. Sanders). 
\end{acknowledgements}
%
%
\bibliographystyle{aa}
\bibliography{biblio}

\begin{thebibliography}{99}
\expandafter\ifx\csname natexlab\endcsname\relax\def\natexlab#1{#1}\fi

\bibitem[{Abolmasov {et~al.}(2009)Abolmasov, Karpov, \&
  Kotani}]{abolmasov_optically_2009}
Abolmasov, P., Karpov, S., \& Kotani, T. 2009, Publications of the Astronomical
  Society of Japan, 61, 213

\bibitem[{Abolmasov {et~al.}(2007)Abolmasov, Swartz, Fabrika, Ghosh,
  Sholukhova, \& Tennant}]{abolmasov_optical_2007}
Abolmasov, P.~K., Swartz, D.~A., Fabrika, S., {et~al.} 2007, ApJ, 668, 124

\bibitem[{Alston {et~al.}(2021)Alston, Pinto, Barret, D'Ai, Del~Santo,
  Earnshaw, Fabian, Fuerst, Kara, Kosec, Middleton, Parker, Pintore, Robba,
  Roberts, Sathyaprakash, Walton, \& Ambrosi}]{alston_quasi-periodic_2021}
Alston, W.~N., Pinto, C., Barret, D., {et~al.} 2021, arXiv:2104.11163
  [astro-ph], arXiv: 2104.11163

\bibitem[{An {et~al.}(2016)An, Lu, \& Wang}]{an_temporal_2016}
An, T., Lu, X.-L., \& Wang, J.-Y. 2016, A\&A, 585, A89, publisher: EDP Sciences

\bibitem[{Arnaud(1996)}]{arnaud_xspec:_1996}
Arnaud, K.~A. 1996, Astronomical Data Analysis Software and Systems V, 101, 17

\bibitem[{Bachetti {et~al.}(2014)Bachetti, Harrison, Walton, Grefenstette,
  Chakrabarty, Fürst, Barret, Beloborodov, Boggs, Christensen, Craig, Fabian,
  Hailey, Hornschemeier, Kaspi, Kulkarni, Maccarone, Miller, Rana, Stern,
  Tendulkar, Tomsick, Webb, \& Zhang}]{bachetti_ultraluminous_2014}
Bachetti, M., Harrison, F.~A., Walton, D.~J., {et~al.} 2014, Nature, 514, 202

\bibitem[{Bachetti {et~al.}(2020)Bachetti, Maccarone, Brightman, Brumback,
  Fürst, Harrison, Heida, Israel, Middleton, Tomsick, Webb, \&
  Walton}]{bachetti_all_2020}
Bachetti, M., Maccarone, T.~J., Brightman, M., {et~al.} 2020, ApJ, 891, 44,
  publisher: American Astronomical Society

\bibitem[{Bachetti {et~al.}(2013)Bachetti, Rana, Walton, Barret, Harrison,
  Boggs, Christensen, Craig, Fabian, Fürst, Grefenstette, Hailey,
  Hornschemeier, Madsen, Miller, Ptak, Stern, Webb, \&
  Zhang}]{bachetti_ultraluminous_2013}
Bachetti, M., Rana, V., Walton, D.~J., {et~al.} 2013, ApJ, 778, 163

\bibitem[{Belloni(2010)}]{belloni_states_2010}
Belloni, T.~M. 2010, 794, 53

\bibitem[{Brightman {et~al.}(2019)Brightman, Harrison, Bachetti, Xu, Fürst,
  Walton, Ptak, Yukita, \& Zezas}]{brightman_60_2019}
Brightman, M., Harrison, F.~A., Bachetti, M., {et~al.} 2019, The Astrophysical
  Journal, 873, 115

\bibitem[{Burrows {et~al.}(2005)Burrows, Hill, Nousek, Kennea, Wells, Osborne,
  Abbey, Beardmore, Mukerjee, Short, Chincarini, Campana, Citterio, Moretti,
  Pagani, Tagliaferri, Giommi, Capalbi, Tamburelli, Angelini, Cusumano,
  Bräuninger, Burkert, \& Hartner}]{burrows_swift_2005}
Burrows, D.~N., Hill, J.~E., Nousek, J.~A., {et~al.} 2005, Space Science
  Reviews, 120, 165

\bibitem[{Cardelli {et~al.}(1989)Cardelli, Clayton, \&
  Mathis}]{cardelli_relationship_1989}
Cardelli, J.~A., Clayton, G.~C., \& Mathis, J.~S. 1989, The Astrophysical
  Journal, 345, 245

\bibitem[{Carpano {et~al.}(2018)Carpano, Haberl, Maitra, \&
  Vasilopoulos}]{carpano_discovery_2018}
Carpano, S., Haberl, F., Maitra, C., \& Vasilopoulos, G. 2018, Monthly Notices
  of the Royal Astronomical Society: Letters, 476, L45

\bibitem[{Cash(1979)}]{cash_parameter_1979}
Cash, W. 1979, The Astrophysical Journal, 228, 939

\bibitem[{Corbet \& Krimm(2013)}]{corbet_superorbital_2013}
Corbet, R. H.~D. \& Krimm, H.~A. 2013, The Astrophysical Journal, 778, 45

\bibitem[{Cseh {et~al.}(2014)Cseh, Kaaret, Corbel, Grisé, Lang, Körding,
  Falcke, Jonker, Miller-Jones, Farrell, Yang, Paragi, \&
  Frey}]{cseh_unveiling_2014}
Cseh, D., Kaaret, P., Corbel, S., {et~al.} 2014, Mon Not R Astron Soc Lett,
  439, L1

\bibitem[{Dauser {et~al.}(2017)Dauser, Middleton, \&
  Wilms}]{dauser_modelling_2017}
Dauser, T., Middleton, M., \& Wilms, J. 2017, Mon Not R Astron Soc, 466, 2236,
  publisher: Oxford Academic

\bibitem[{Done \& Gierliński(2003)}]{done_observing_2003}
Done, C. \& Gierliński, M. 2003, Mon Not R Astron Soc, 342, 1041

\bibitem[{Evans {et~al.}(2009)Evans, Beardmore, Page, Osborne, O'Brien,
  Willingale, Starling, Burrows, Godet, Vetere, Racusin, Goad, Wiersema,
  Angelini, Capalbi, Chincarini, Gehrels, Kennea, Margutti, Morris, Mountford,
  Pagani, Perri, Romano, \& Tanvir}]{evans_methods_2009}
Evans, P.~A., Beardmore, A.~P., Page, K.~L., {et~al.} 2009, Mon Not R Astron
  Soc, 397, 1177, publisher: Oxford Academic

\bibitem[{Evans {et~al.}(2007)Evans, Beardmore, Page, Tyler, Osborne, Goad,
  O'Brien, Vetere, Racusin, Morris, Burrows, Capalbi, Perri, Gehrels, \&
  Romano}]{evans_online_2007}
Evans, P.~A., Beardmore, A.~P., Page, K.~L., {et~al.} 2007, A\&A, 469, 379,
  number: 1 Publisher: EDP Sciences

\bibitem[{Fabrika {et~al.}(2015)Fabrika, Ueda, Vinokurov, Sholukhova, \&
  Shidatsu}]{fabrika_supercritical_2015}
Fabrika, S., Ueda, Y., Vinokurov, A., Sholukhova, O., \& Shidatsu, M. 2015,
  Nature Phys, 11, 551, number: 7 Publisher: Nature Publishing Group

\bibitem[{Farrell {et~al.}(2009)Farrell, Webb, Barret, Godet, \&
  Rodrigues}]{farrell_intermediate-mass_2009}
Farrell, S.~A., Webb, N.~A., Barret, D., Godet, O., \& Rodrigues, J.~M. 2009,
  Nature, 460, 73

\bibitem[{Fender {et~al.}(2004)Fender, Belloni, \& Gallo}]{fender_towards_2004}
Fender, R.~P., Belloni, T.~M., \& Gallo, E. 2004, Monthly Notices of the Royal
  Astronomical Society, 355, 1105

\bibitem[{Feng {et~al.}(2016)Feng, Tao, Kaaret, \& Grisé}]{feng_nature_2016}
Feng, H., Tao, L., Kaaret, P., \& Grisé, F. 2016, ApJ, 831, 117

\bibitem[{Foreman-Mackey {et~al.}(2017)Foreman-Mackey, Agol, Ambikasaran, \&
  Angus}]{foreman-mackey_fast_2017}
Foreman-Mackey, D., Agol, E., Ambikasaran, S., \& Angus, R. 2017, AJ, 154, 220,
  publisher: American Astronomical Society

\bibitem[{Fürst {et~al.}(2016)Fürst, Walton, Harrison, Stern, Barret,
  Brightman, Fabian, Madsen, Middleton, Miller, Pottschmidt, Ptak, \&
  Rana}]{furst_discovery_2016}
Fürst, F., Walton, D., Harrison, F., {et~al.} 2016, The Astrophysical Journal,
  831

\bibitem[{Fürst {et~al.}(2018)Fürst, Walton, Heida, Harrison, Barret,
  Brightman, Fabian, Middleton, Pinto, Rana, Tramper, Webb, \&
  Kretschmar}]{furst_tale_2018}
Fürst, F., Walton, D.~J., Heida, M., {et~al.} 2018, Astronomy \& Astrophysics,
  616, A186

\bibitem[{Gladstone {et~al.}(2007)Gladstone, Done, \&
  Gierlinski}]{gladstone_analysing_2007}
Gladstone, J., Done, C., \& Gierlinski, M. 2007, Monthly Notices of the Royal
  Astronomical Society, 378, 13

\bibitem[{Gladstone {et~al.}(2009)Gladstone, Roberts, \&
  Done}]{gladstone_ultraluminous_2009}
Gladstone, J.~C., Roberts, T.~P., \& Done, C. 2009, Mon Not R Astron Soc, 397,
  1836

\bibitem[{Godet {et~al.}(2009)Godet, Barret, Webb, Farrell, \&
  Gehrels}]{godet_first_2009}
Godet, O., Barret, D., Webb, N.~A., Farrell, S.~A., \& Gehrels, N. 2009, ApJ,
  705, L109

\bibitem[{Grisé {et~al.}(2013)Grisé, Kaaret, Corbel, Cseh, \&
  Feng}]{grise_long-term_2013}
Grisé, F., Kaaret, P., Corbel, S., Cseh, D., \& Feng, H. 2013, Mon Not R
  Astron Soc, 433, 1023

\bibitem[{Grisé {et~al.}(2010)Grisé, Kaaret, Feng, Kajava, \&
  Farrell}]{grise_x-ray_2010}
Grisé, F., Kaaret, P., Feng, H., Kajava, J. J.~E., \& Farrell, S.~A. 2010, The
  Astrophysical Journal, 724, L148

\bibitem[{Guo {et~al.}(2019)Guo, Sun, Gu, \& Yi}]{guo_thick-disc_2019}
Guo, J., Sun, M., Gu, W.-M., \& Yi, T. 2019, Monthly Notices of the Royal
  Astronomical Society, 485, 2558

\bibitem[{Gúrpide {et~al.}(2021)Gúrpide, Godet, Koliopanos, Webb, \&
  Olive}]{gurpide_long-term_2021}
Gúrpide, A., Godet, O., Koliopanos, F., Webb, N., \& Olive, J.-F. 2021, A\&A,
  649, A104, publisher: EDP Sciences

\bibitem[{Harrison {et~al.}(2013)Harrison, Craig, Christensen, Hailey, Zhang,
  Boggs, Stern, Cook, Forster, Giommi, Grefenstette, Kim, Kitaguchi, Koglin,
  Madsen, Mao, Miyasaka, Mori, Perri, Pivovaroff, Puccetti, Rana, Westergaard,
  Willis, Zoglauer, An, Bachetti, Barrière, Bellm, Bhalerao, Brejnholt,
  Fuerst, Liebe, Markwardt, Nynka, Vogel, Walton, Wik, Alexander, Cominsky,
  Hornschemeier, Hornstrup, Kaspi, Madejski, Matt, Molendi, Smith, Tomsick,
  Ajello, Ballantyne, Baloković, Barret, Bauer, Blandford, Brandt, Brenneman,
  Chiang, Chakrabarty, Chenevez, Comastri, Dufour, Elvis, Fabian, Farrah,
  Fryer, Gotthelf, Grindlay, Helfand, Krivonos, Meier, Miller, Natalucci, Ogle,
  Ofek, Ptak, Reynolds, Rigby, Tagliaferri, Thorsett, Treister, \&
  Urry}]{harrison_nuclear_2013}
Harrison, F.~A., Craig, W.~W., Christensen, F.~E., {et~al.} 2013, The
  Astrophysical Journal, 770, 103

\bibitem[{Heida {et~al.}(2019)Heida, Lau, Davies, Brightman, Fürst,
  Grefenstette, Kennea, Tramper, Walton, \& Harrison}]{heida_discovery_2019}
Heida, M., Lau, R.~M., Davies, B., {et~al.} 2019, ApJL, 883, L34, publisher:
  IOP Publishing

\bibitem[{Heil {et~al.}(2009)Heil, Vaughan, \& Roberts}]{heil_systematic_2009}
Heil, L.~M., Vaughan, S., \& Roberts, T.~P. 2009, Mon Not R Astron Soc, 397,
  1061

\bibitem[{{HI4PI Collaboration} {et~al.}(2016){HI4PI Collaboration}, Bekhti,
  Flöer, Keller, Kerp, Lenz, Winkel, Bailin, Calabretta, Dedes, Ford, Gibson,
  Haud, Janowiecki, Kalberla, Lockman, McClure-Griffiths, Murphy, Nakanishi,
  Pisano, \& Staveley-Smith}]{hi4pi_collaboration_hi4pi_2016}
{HI4PI Collaboration}, Bekhti, N.~B., Flöer, L., {et~al.} 2016, A\&A, 594,
  A116, publisher: EDP Sciences

\bibitem[{Horne \& Baliunas(1986)}]{horne_prescription_1986}
Horne, J.~H. \& Baliunas, S.~L. 1986, The Astrophysical Journal, 302, 757

\bibitem[{Israel {et~al.}(2017)Israel, Belfiore, Stella, Esposito, Casella,
  De~Luca, Marelli, Papitto, Perri, Puccetti, Castillo, Salvetti, Tiengo,
  Zampieri, D'Agostino, Greiner, Haberl, Novara, Salvaterra, Turolla, Watson,
  Wilms, \& Wolter}]{israel_accreting_2017}
Israel, G.~L., Belfiore, A., Stella, L., {et~al.} 2017, Science, 355, 817

\bibitem[{Jansen {et~al.}(2001)Jansen, Lumb, Altieri, Clavel, Ehle, Erd,
  Gabriel, Guainazzi, Gondoin, Much, Munoz, Santos, Schartel, Texier, \&
  Vacanti}]{jansen_xmm-newton_2001}
Jansen, F., Lumb, D., Altieri, B., {et~al.} 2001, Astronomy \& Astrophysics,
  365, L1

\bibitem[{Kaaret \& Feng(2009)}]{kaaret_x-ray_2009}
Kaaret, P. \& Feng, H. 2009, ApJ, 702, 1679, publisher: IOP Publishing

\bibitem[{Kaaret {et~al.}(2017)Kaaret, Feng, \&
  Roberts}]{kaaret_ultraluminous_2017}
Kaaret, P., Feng, H., \& Roberts, T.~P. 2017, Annual Review of Astronomy and
  Astrophysics, 55, 303, \_eprint:
  https://doi.org/10.1146/annurev-astro-091916-055259

\bibitem[{Kalberla {et~al.}(2005)Kalberla, Burton, Hartmann, Arnal, Bajaja,
  Morras, \& Pöppel}]{kalberla_leiden/argentine/bonn_2005}
Kalberla, P. M.~W., Burton, W.~B., Hartmann, D., {et~al.} 2005, Astronomy and
  Astrophysics, 440, 775

\bibitem[{Kawashima {et~al.}(2012)Kawashima, Ohsuga, Mineshige, Yoshida,
  Heinzeller, \& Matsumoto}]{kawashima_comptonized_2012}
Kawashima, T., Ohsuga, K., Mineshige, S., {et~al.} 2012, ApJ, 752, 18

\bibitem[{King(2009)}]{king_masses_2009}
King, A.~R. 2009, Mon Not R Astron Soc Lett, 393, L41

\bibitem[{Kong {et~al.}(2016)Kong, Hu, Lin, Li, Jin, Liu, \&
  Yen}]{kong_possible_2016}
Kong, A. K.~H., Hu, C.-P., Lin, L. C.-C., {et~al.} 2016, Monthly Notices of the
  Royal Astronomical Society, 461, 4395

\bibitem[{Kozai(1962)}]{kozai_secular_1962}
Kozai, Y. 1962, The Astronomical Journal, 67, 591

\bibitem[{Lau {et~al.}(2019)Lau, Heida, Walton, Kasliwal, Adams, Cody, De,
  Gehrz, Fürst, Jencson, Kennea, \& Masci}]{lau_uncovering_2019}
Lau, R.~M., Heida, M., Walton, D.~J., {et~al.} 2019, ApJ, 878, 71, publisher:
  American Astronomical Society

\bibitem[{Liu(2008)}]{liu_no_2008}
Liu, J.-F. 2008, The Astrophysical Journal Supplement Series, 177, 181

\bibitem[{Liu {et~al.}(2004)Liu, Bregman, \& Seitzer}]{liu_optical_2004}
Liu, J.-F., Bregman, J.~N., \& Seitzer, P. 2004, ApJ, 602, 249, publisher: IOP
  Publishing

\bibitem[{Lomb(1976)}]{lomb_least-squares_1976}
Lomb, N.~R. 1976, Astrophysics and Space Science, 39, 447

\bibitem[{Mezcua(2017)}]{mezcua_observational_2017}
Mezcua, M. 2017, Int. J. Mod. Phys. D, 26, 1730021

\bibitem[{Middleton {et~al.}(2018)Middleton, Fragile, Bachetti, Brightman,
  Jiang, Ho, Roberts, Ingram, Dauser, Pinto, Walton, Fuerst, Fabian, \&
  Gehrels}]{middleton_lense-thirring_2018}
Middleton, M.~J., Fragile, P.~C., Bachetti, M., {et~al.} 2018, Mon Not R Astron
  Soc, 475, 154, publisher: Oxford Academic

\bibitem[{Middleton {et~al.}(2019)Middleton, Fragile, Ingram, \&
  Roberts}]{middleton_lensethirring_2019}
Middleton, M.~J., Fragile, P.~C., Ingram, A., \& Roberts, T.~P. 2019, Mon Not R
  Astron Soc, 489, 282, publisher: Oxford Academic

\bibitem[{Middleton {et~al.}(2015{\natexlab{a}})Middleton, Heil, Pintore,
  Walton, \& Roberts}]{middleton_spectral-timing_2015}
Middleton, M.~J., Heil, L., Pintore, F., Walton, D.~J., \& Roberts, T.~P.
  2015{\natexlab{a}}, Monthly Notices of the Royal Astronomical Society, 447,
  3243

\bibitem[{Middleton {et~al.}(2011)Middleton, Roberts, Done, \&
  Jackson}]{middleton_challenging_2011}
Middleton, M.~J., Roberts, T.~P., Done, C., \& Jackson, F.~E. 2011, Mon Not R
  Astron Soc, 411, 644, publisher: Oxford Academic

\bibitem[{Middleton {et~al.}(2015{\natexlab{b}})Middleton, Walton, Fabian,
  Roberts, Heil, Pinto, Anderson, \& Sutton}]{middleton_diagnosing_2015}
Middleton, M.~J., Walton, D.~J., Fabian, A., {et~al.} 2015{\natexlab{b}},
  Monthly Notices of the Royal Astronomical Society, 454, 3134

\bibitem[{Mondal {et~al.}(2021)Mondal, Rozanska, Baginska, Markowitz, \&
  De~Marco}]{mondal_spectral_2021}
Mondal, S., Rozanska, A., Baginska, P., Markowitz, A., \& De~Marco, B. 2021,
  arXiv e-prints, 2104, arXiv:2104.12894

\bibitem[{Mudelsee(2002)}]{mudelsee_tauest_2002}
Mudelsee, M. 2002, Computers \& Geosciences, 28, 69

\bibitem[{Narayan {et~al.}(2017)Narayan, Sa̧dowski, \&
  Soria}]{narayan_spectra_2017}
Narayan, R., Sa̧dowski, A., \& Soria, R. 2017, Mon Not R Astron Soc, 469,
  2997, publisher: Oxford Academic

\bibitem[{Ohsuga \& Mineshige(2007)}]{ohsuga_why_2007}
Ohsuga, K. \& Mineshige, S. 2007, ApJ, 670, 1283, publisher: IOP Publishing

\bibitem[{Pakull \& Mirioni(2002)}]{pakull_optical_2002}
Pakull, M.~W. \& Mirioni, L. 2002, arXiv:astro-ph/0202488, arXiv:
  astro-ph/0202488

\bibitem[{Pinto {et~al.}(2017)Pinto, Alston, Soria, Middleton, Walton, Sutton,
  Fabian, Earnshaw, Urquhart, Kara, \& Roberts}]{pinto_ultraluminous_2017}
Pinto, C., Alston, W., Soria, R., {et~al.} 2017, Mon Not R Astron Soc, 468,
  2865, publisher: Oxford Academic

\bibitem[{Pinto {et~al.}(2020)Pinto, Mehdipour, Walton, Middleton, Roberts,
  Fabian, Guainazzi, Soria, Kosec, \& Ness}]{pinto_thermal_2020}
Pinto, C., Mehdipour, M., Walton, D.~J., {et~al.} 2020, Mon Not R Astron Soc,
  491, 5702, publisher: Oxford Academic

\bibitem[{Pinto {et~al.}(2021)Pinto, Soria, Walton, D'Ai, Pintore, Kosec,
  Alston, Fuerst, Middleton, Roberts, Del~Santo, Barret, Ambrosi, Robba,
  Earnshaw, \& Fabian}]{pinto_xmm-newton_2021}
Pinto, C., Soria, R., Walton, D., {et~al.} 2021, arXiv:2104.11164 [astro-ph],
  arXiv: 2104.11164

\bibitem[{Poutanen {et~al.}(2007)Poutanen, Lipunova, Fabrika, Butkevich, \&
  Abolmasov}]{poutanen_supercritically_2007}
Poutanen, J., Lipunova, G., Fabrika, S., Butkevich, A.~G., \& Abolmasov, P.
  2007, Monthly Notices of the Royal Astronomical Society, 377, 1187

\bibitem[{Quintin {et~al.}(2021)Quintin, Webb, Gúrpide, Bachetti, \&
  Fürst}]{quintin_new_2021}
Quintin, E., Webb, N., Gúrpide, A., Bachetti, M., \& Fürst, F. 2021,
  arXiv:2103.11650 [astro-ph], arXiv: 2103.11650

\bibitem[{Ray {et~al.}(2019)Ray, Guillot, Ho, Kerr, Enoto, Gendreau,
  Arzoumanian, Altamirano, Bogdanov, Campion, Chakrabarty, Deneva, Jaisawal,
  Kozon, Malacaria, Strohmayer, \& Wolff}]{ray_anti-glitches_2019}
Ray, P.~S., Guillot, S., Ho, W. C.~G., {et~al.} 2019, ApJ, 879, 130

\bibitem[{Remillard \& McClintock(2006)}]{remillard_x-ray_2006}
Remillard, R.~A. \& McClintock, J.~E. 2006, Annual Review of Astronomy and
  Astrophysics, 44, 49

\bibitem[{Roberts {et~al.}(2006)Roberts, Kilgard, Warwick, Goad, \&
  Ward}]{roberts_chandra_2006}
Roberts, T.~P., Kilgard, R.~E., Warwick, R.~S., Goad, M.~R., \& Ward, M.~J.
  2006, Monthly Notices of the Royal Astronomical Society, 371, 1877

\bibitem[{Robinson(1977)}]{robinson_estimation_1977}
Robinson, P.~M. 1977, Stochastic Processes and their Applications, 6, 9

\bibitem[{Rodríguez-Castillo {et~al.}(2020)Rodríguez-Castillo, Israel,
  Belfiore, Bernardini, Esposito, Pintore, Luca, Papitto, Stella, Tiengo,
  Zampieri, Bachetti, Brightman, Casella, D'Agostino, Dall'Osso, Earnshaw,
  Fürst, Haberl, Harrison, Mapelli, Marelli, Middleton, Pinto, Roberts,
  Salvaterra, Turolla, Walton, \& Wolter}]{rodriguez-castillo_discovery_2020}
Rodríguez-Castillo, G.~A., Israel, G.~L., Belfiore, A., {et~al.} 2020, ApJ,
  895, 60, publisher: American Astronomical Society

\bibitem[{Sathyaprakash {et~al.}(2019)Sathyaprakash, Roberts, Walton, Fuerst,
  Bachetti, Pinto, Alston, Earnshaw, Fabian, Middleton, \&
  Soria}]{sathyaprakash_discovery_2019}
Sathyaprakash, R., Roberts, T.~P., Walton, D.~J., {et~al.} 2019, Monthly
  Notices of the Royal Astronomical Society, 488, L35

\bibitem[{Scargle(1981)}]{scargle_studies_1981}
Scargle, J.~D. 1981, The Astrophysical Journal Supplement Series, 45, 1

\bibitem[{Scargle(1982)}]{scargle_studies_1982}
Scargle, J.~D. 1982, The Astrophysical Journal, 263, 835

\bibitem[{Schulz \& Mudelsee(2002)}]{schulz_redfit_2002}
Schulz, M. \& Mudelsee, M. 2002, Computers \& Geosciences, 28, 421

\bibitem[{Schulz \& Stattegger(1997)}]{schulz_spectrum_1997}
Schulz, M. \& Stattegger, K. 1997, Computers \& Geosciences, 23, 929

\bibitem[{Schwarz(1978)}]{schwarz_estimating_1978}
Schwarz, G. 1978, Ann. Statist., 6, 461, publisher: Institute of Mathematical
  Statistics

\bibitem[{Servillat {et~al.}(2011)Servillat, Farrell, Lin, Godet, Barret, \&
  Webb}]{servillat_x-ray_2011}
Servillat, M., Farrell, S.~A., Lin, D., {et~al.} 2011, ApJ, 743, 6

\bibitem[{Shakura \& Sunyaev(1973)}]{shakura_black_1973}
Shakura, N.~I. \& Sunyaev, R.~A. 1973, Astronomy and Astrophysics, 24, 337

\bibitem[{Stobbart {et~al.}(2006)Stobbart, Roberts, \&
  Wilms}]{stobbart_xmmnewton_2006}
Stobbart, A.-M., Roberts, T.~P., \& Wilms, J. 2006, Mon Not R Astron Soc, 368,
  397

\bibitem[{Strüder {et~al.}(2001)Strüder, Briel, Dennerl, Hartmann,
  Kendziorra, Meidinger, Pfeffermann, Reppin, Aschenbach, Bornemann,
  Bräuninger, Burkert, Elender, Freyberg, Haberl, Hartner, Heuschmann,
  Hippmann, Kastelic, Kemmer, Kettenring, Kink, Krause, Müller, Oppitz,
  Pietsch, Popp, Predehl, Read, Stephan, Stötter, Trümper, Holl, Kemmer,
  Soltau, Stötter, Weber, Weichert, von Zanthier, Carathanassis, Lutz,
  Richter, Solc, Böttcher, Kuster, Staubert, Abbey, Holland, Turner, Balasini,
  Bignami, La~Palombara, Villa, Buttler, Gianini, Lainé, Lumb, \&
  Dhez}]{struder_european_2001}
Strüder, L., Briel, U., Dennerl, K., {et~al.} 2001, Astronomy \& Astrophysics,
  365, L18

\bibitem[{Sutton {et~al.}(2013)Sutton, Roberts, \&
  Middleton}]{sutton_ultraluminous_2013}
Sutton, A.~D., Roberts, T.~P., \& Middleton, M.~J. 2013, Monthly Notices of the
  Royal Astronomical Society, 435, 1758

\bibitem[{Takeuchi {et~al.}(2013)Takeuchi, Ohsuga, \&
  Mineshige}]{takeuchi_clumpy_2013}
Takeuchi, S., Ohsuga, K., \& Mineshige, S. 2013, Publ Astron Soc Jpn Nihon
  Tenmon Gakkai, 65, publisher: Oxford Academic

\bibitem[{Tao {et~al.}(2012)Tao, Kaaret, Feng, \& Grisé}]{tao_nature_2012}
Tao, L., Kaaret, P., Feng, H., \& Grisé, F. 2012, ApJ, 750, 110, publisher:
  American Astronomical Society

\bibitem[{Townsend \& Charles(2020)}]{townsend_orbital_2020}
Townsend, L.~J. \& Charles, P.~A. 2020, Monthly Notices of the Royal
  Astronomical Society: Letters, 495, L139

\bibitem[{Tsygankov {et~al.}(2016)Tsygankov, Mushtukov, Suleimanov, \&
  Poutanen}]{tsygankov_propeller_2016}
Tsygankov, S.~S., Mushtukov, A.~A., Suleimanov, V.~F., \& Poutanen, J. 2016,
  Mon Not R Astron Soc, 457, 1101

\bibitem[{Turner {et~al.}(2001)Turner, Abbey, Arnaud, Balasini, Barbera,
  Belsole, Bennie, Bernard, Bignami, Boer, Briel, Butler, Cara, Chabaud, Cole,
  Collura, Conte, Cros, Denby, Dhez, Di~Coco, Dowson, Ferrando, Ghizzardi,
  Gianotti, Goodall, Gretton, Griffiths, Hainaut, Hochedez, Holland, Jourdain,
  Kendziorra, Lagostina, Laine, La~Palombara, Lortholary, Lumb, Marty, Molendi,
  Pigot, Poindron, Pounds, Reeves, Reppin, Rothenflug, Salvetat, Sauvageot,
  Schmitt, Sembay, Short, Spragg, Stephen, Strüder, Tiengo, Trifoglio,
  Trümper, Vercellone, Vigroux, Villa, Ward, Whitehead, \&
  Zonca}]{turner_european_2001}
Turner, M. J.~L., Abbey, A., Arnaud, M., {et~al.} 2001, A\&A, 365, L27

\bibitem[{Urquhart \& Soria(2016)}]{urquhart_optically_2016}
Urquhart, R. \& Soria, R. 2016, MNRAS, 456, 1859

\bibitem[{VanderPlas(2018)}]{vanderplas_understanding_2018}
VanderPlas, J.~T. 2018, ApJS, 236, 16, publisher: American Astronomical Society

\bibitem[{Vasilopoulos {et~al.}(2018)Vasilopoulos, Haberl, Carpano, \&
  Maitra}]{vasilopoulos_ngc_2018}
Vasilopoulos, G., Haberl, F., Carpano, S., \& Maitra, C. 2018, A\&A, 620, L12,
  publisher: EDP Sciences

\bibitem[{Vasilopoulos {et~al.}(2020)Vasilopoulos, Lander, Koliopanos, \&
  Bailyn}]{vasilopoulos_m51_2020}
Vasilopoulos, G., Lander, S.~K., Koliopanos, F., \& Bailyn, C.~D. 2020, Mon Not
  R Astron Soc, 491, 4949, publisher: Oxford Academic

\bibitem[{Vasilopoulos {et~al.}(2019)Vasilopoulos, Petropoulou, Koliopanos,
  Ray, Bailyn, Haberl, \& Gendreau}]{vasilopoulos_ngc_2019}
Vasilopoulos, G., Petropoulou, M., Koliopanos, F., {et~al.} 2019, Mon Not R
  Astron Soc, 488, 5225, publisher: Oxford Academic

\bibitem[{Vaughan {et~al.}(2016)Vaughan, Uttley, Markowitz, Huppenkothen,
  Middleton, Alston, Scargle, \& Farr}]{vaughan_false_2016}
Vaughan, S., Uttley, P., Markowitz, A.~G., {et~al.} 2016, Monthly Notices of
  the Royal Astronomical Society, 461, 3145

\bibitem[{Vinokurov {et~al.}(2013)Vinokurov, Fabrika, \&
  Atapin}]{vinokurov_ultra-luminous_2013}
Vinokurov, A., Fabrika, S., \& Atapin, K. 2013, Astrophys. Bull., 68, 139

\bibitem[{Weisskopf {et~al.}(2000)Weisskopf, Tananbaum, Van~Speybroeck, \&
  O'Dell}]{weisskopf_chandra_2000}
Weisskopf, M.~C., Tananbaum, H.~D., Van~Speybroeck, L.~P., \& O'Dell, S.~L.
  2000, 4012, 2, conference Name: X-Ray Optics, Instruments, and Missions III
  Place: eprint: arXiv:astro-ph/0004127

\bibitem[{Zdziarski {et~al.}(2007)Zdziarski, Wen, \&
  Gierliński}]{zdziarski_superorbital_2007}
Zdziarski, A.~A., Wen, L., \& Gierliński, M. 2007, Monthly Notices of the
  Royal Astronomical Society, 377, 1006

\bibitem[{Zhang {et~al.}(2020)Zhang, Altamirano, Cúneo, Alabarta, Enoto,
  Homan, Remillard, Uttley, Vincentelli, Arzoumanian, Bult, Gendreau,
  Markwardt, Sanna, Strohmayer, Steiner, Basak, Neilsen, \&
  Tombesi}]{zhang_nicer_2020}
Zhang, L., Altamirano, D., Cúneo, V.~A., {et~al.} 2020, Monthly Notices of the
  Royal Astronomical Society, 499, 851

\end{thebibliography}

\end{document}